\documentclass[%
reprint,
 amsmath,amssymb,
 aps,
onecolumn
]{revtex4-2}

\usepackage{dcolumn}
\usepackage{bm}

\usepackage{float}
\usepackage{array}
\usepackage[english]{babel}
\usepackage[utf8]{inputenc}
\usepackage{amsmath}
\usepackage{amsfonts}
\usepackage{graphicx}
\usepackage{dsfont}
\usepackage{ulem} 
\usepackage{algorithm}
\usepackage[noend]{algpseudocode}
\usepackage{bbold}
\usepackage{mathtools}
\usepackage{amssymb}
\usepackage{bm}
\usepackage{bbm}
\usepackage{tabularx, booktabs}
\usepackage{xspace}
\usepackage{listings}
\usepackage{lipsum}
\usepackage{chemformula}
\usepackage{tikz}
\usepackage{physics}
\usepackage{mathtools}
\usepackage{appendix}
\usepackage{mathrsfs}

\newcommand{\be}{\begin{eqnarray}}
\newcommand{\ee}{\end{eqnarray}}

\newcommand{\bv}{{\mathbf{v}}}

\newcommand{\ie}{{\it{i.e.}}}

\begin{document}

\title{Local Potential Functional Embedding Theory: A Self-Consistent Flavor of Density Functional Theory for Lattices without Density Functionals}

\author{Sajanthan Sekaran$^1$}
 \email{s.sekaran@unistra.fr}
\author{Matthieu Sauban\`{e}re$^2$}
    \email{matthieu.saubanere@umontpellier.fr}
\author{Emmanuel Fromager$^1$}
 \email{fromagere@unistra.fr}
\affiliation{%
 $^1$Laboratoire de Chimie Quantique, Institut de Chimie,\\
 CNRS/Université de Strasbourg, 4 rue Blaise Pascal, 67000 Strasbourg, France\\
 $^2$ ICGM, Universit\'{e} de Montpellier,\\
CNRS, ENSCM, Montpellier, France
}%

\begin{abstract}
The recently proposed {Householder transformed density-matrix functional embedding theory} (Ht-DMFET) [Sekaran {\it et al.}, Phys. Rev. B 104, 035121 (2021)], which is equivalent to (but formally simpler than) {density matrix embedding theory} (DMET) in the  non-interacting case, is revisited from the perspective of density-functional theory (DFT). An in-principle-exact density-functional version of Ht-DMFET is derived for the one-dimensional Hubbard lattice with a single embedded impurity. On the basis of well-identified density-functional approximations, a local potential functional embedding theory (LPFET) is formulated and implemented. Even though LPFET performs better than Ht-DMFET in the low-density regime, in particular when electron correlation is strong, both methods are unable to describe the density-driven Mott--Hubbard transition, as expected. These results combined with our formally exact density-functional embedding theory reveal that a single statically embedded impurity can in principle describe the gap opening, provided that the complementary correlation potential (that describes the interaction of the embedding cluster with its environment, which is simply neglected in both Ht-DMFET and LPFET) exhibits a derivative discontinuity (DD) at half filling. The extension of LPFET to multiple impurities (which would enable to circumvent the modeling of DDs) and its generalization to quantum chemical Hamiltonians are left for future work.
\end{abstract}


\maketitle

\section{Introduction}

Kohn--Sham density-functional theory (KS-DFT)~\cite{KStheory_1965} has become over the last two decades the method of choice for computational chemistry and physics studies, essentially because it often provides a relatively accurate description of the electronic structure of large molecular or extended systems at a low computational cost. The major simplification of the electronic structure problem in KS-DFT lies in the fact that the ground-state energy is evaluated, in principle exactly, from a non-interacting single-configuration wave function, which is simply referred to as the KS determinant. The latter is obviously not the exact solution to the Schr\"{o}dinger equation. However, its density matches the exact interacting ground-state density, so that the Hartree-exchange-correlation (Hxc) energy of the physical system, which is induced by the electronic repulsion, can be recovered from an appropriate (in principle exact and universal) Hxc density functional. Despite the success of KS-DFT, standard density-functional approximations still fail in describing strongly correlated electrons. To overcome this issue, various strategies have been explored and improved over the years, both in condensed matter physics~\cite{Anisimov_1997_lda_plus_U,Anisimov_1997,PRB98_Lichtenstein_LDA_plus_DMFT,kotliar2006reviewDMFT,Haule_2ble_counting_DMFT-DFT_2015,requist2019model} and quantum chemistry~\cite{CR18_Truhlar_Multiconf_DFTs}. Note that, in the latter case, in-principle-exact multi-determinantal extensions of DFT based on the adiabatic connection formalism have been developed~\cite{savinbook,toulouse2004long,sharkas2011double,fromager2015exact}. In these approaches, the KS system is only referred to in the design of density-functional approximations. In practice, a single (partially-interacting) many-body wave function is calculated self-consistently and the complement to the partial interaction energy is described with an appropriate density functional (which differs from the conventional xc one). In other words, there is no KS construction in the actual calculation. Some of these concepts have been reused in the study of model lattice Hamiltonians~\cite{fromager2015exact,senjean2018site}. A similar strategy will be adopted in the present work, with an important difference though. The {\it reduced-in-size} correlated density-functional many-body wave function that we will introduce will be extracted from a quantum embedding theory where the KS determinant of the full system is a key ingredient that must be evaluated explicitly.\\ 

Quantum embedding theory~\cite{IJQC20_Adam-Michele_embedding_special_issue} is at first sight a completely different approach to the strong electron correlation problem. Interestingly, some of its implementations, like the \textit{density matrix embedding theory} (DMET)~\cite{knizia2012density,knizia2013density,tsuchimochi2015density,welborn2016bootstrap,sun2016quantum,wouters2016practical,wu2019projected,JCTC20_Chan_ab-initio_DMET,faulstich2022_vrep}, rely on a reference Slater determinant that is computed for the full system. This is also the case in practical embedding calculations based on the exact factorization formalism~\cite{PRL20_Lacombe_embedding_via_EF,PRL21_Requist_EF_electrons}. Unlike the well-established \textit{dynamical mean-field theory} (DMFT)~\cite{georges1992hubbard,georges1996limitdimension,kotliar2004strongly,held2007electronic,zgid2011DMFTquantum}, which relies on the one-electron Green's function, DMET is a static theory of ground electronic states. Most importantly, the bath, in which a fragment of the original system (referred to as impurity when it is a single localized orbital) is embedded, is drastically reduced in size in DMET. As a result, the ``impurity+bath'' embedding cluster can be accurately (if not exactly) described with wave function-based quantum chemical methods. The authors have shown recently that the Schmidt decomposition of the reference Slater determinant, which is central in DMET, can be recast into a (one-electron reduced) density-matrix functional Householder transformation~\cite{sekaran2021}, which is much simpler to implement. The approach, in which the bath orbitals can in principle be correlated directly through the density matrix~\cite{sekaran2021}, is referred to as \textit{ Householder~transformed~density matrix~functional~embedding~theory} (Ht-DMFET). Since the seminal work of Knizia and Chan on DMET~\cite{knizia2012density}, various connections with DMFT and related approaches have been established~\cite{ayral2017dynamical,lee2019rotationally,fertitta2018rigorous,JCP19_Booth_Ew-DMET_hydrogen_chain,   PRB21_Booth_effective_dynamics_static_embedding,PRX21_Lee_SlaveBoson_resp_functions-superconductivity}. Connections with DFT have been less explored, and only at the approximate level of theory. We can refer to the \textit{density embedding theory} (DET) of Bulik {\it et al.}~\cite{bulik2014density}, which is a simplified version of DMET where only the diagonal elements of the embedded density matrix are mapped onto the reference Slater determinant of the full system. More recently, Senjean~\cite{senjean2019projected} combined DFT for lattices~\cite{lima2003density,DFT_ModelHamiltonians} with DMET, and Mordovina {\it et al.}~\cite{mordovina2019self} (see also Ref.~\cite{Theophilou_2021}) proposed a {\it self-consistent density-functional embedding} (SDE), where the KS determinant is explicitly used as the reference wave function in the DMET algorithm.\\

In the present work, an in-principle-exact combination of KS-DFT with DMET is derived for the one-dimensional (1D) Hubbard lattice, as a proof of concept. For that purpose, we use the density-matrix functional Householder transformation introduced recently by the authors~\cite{sekaran2021}. On the basis of well-identified density-functional approximations, we propose and implement a {\it local potential functional embedding theory} (LPFET) where the Hxc potential is evaluated self-consistently in the lattice by ``learning'' from the embedding cluster at each iteration of the optimization process. LPFET can be seen as a flavor of KS-DFT where no density functional is actually used.\\    

The paper is organized as follows. After a short introduction to the 1D Hubbard model in Sec.~\ref{subsec:1D_hub}, a detailed review of Ht-DMFET is presented in Sec.~\ref{subsec:review_Ht-DMFET}, for clarity and completeness. An exact density-functional reformulation of the theory is then proposed in Sec.~\ref{subsec:exact_dfe_dft}. The resulting approximate LPFET and its comparison with SDE are detailed in Secs.~\ref{sec:LPFET} and Sec.~\ref{subsec:sde_comparison}, respectively. The LPFET algorithm is summarized in Sec.~\ref{sec:lpfet_algo}. Results obtained for a 1000-site Hubbard ring are presented and discussed in Sec.~\ref{sec:results}. The conclusion and perspectives are finally given in Sec.~\ref{sec:conclusion}. 

\section{Theory}\label{sec:theory}

\subsection{One-dimensional Hubbard lattice}\label{subsec:1D_hub}

By analogy with Ref.~\cite{sekaran2021}, various quantum embedding strategies will be discussed in the following within the simple but nontrivial uniform 1D Hubbard model. The corresponding lattice Hamiltonian (for a $L$-site ring) reads as 
\be\label{eq:1D_Hubbard_Hamilt}
\hat{H}=\hat{T}+\hat{U}+v_{\rm ext}\hat{N},
\ee
where the hopping operator (written in second quantization), 
\be\label{eq:hopping_operator}
\hat{T}=-t\sum^{L-1}_{i=0}\sum_{\sigma=\uparrow,\downarrow}\left(\hat{c}^\dagger_{i\sigma}\hat{c}_{(i+1)\sigma}
+\hat{c}_{(i+1)\sigma}^\dagger\hat{c}_{i\sigma}
\right),
\ee
with parameter $t$, is the analog for lattices of the kinetic energy operator. For convenience, we will systematically use periodic boundary conditions, \ie, $\hat{c}^\dagger_{L\sigma}\equiv \hat{c}^\dagger_{0\sigma}$. On-site repulsions only are taken into account in the two-electron repulsion operator $\hat{U}$, \ie,
\be
\hat{U}=\sum^{L-1}_{i=0}\hat{U}_i,
\ee
where $\hat{U}_i=U\hat{n}_{i\uparrow}\hat{n}_{i\downarrow}$, $U$ is the parameter that controls the strength of the interaction, and $\hat{n}_{i\sigma}=\hat{c}_{i\sigma}^\dagger\hat{c}_{i\sigma}$ is a site occupation operator for spin $\sigma$. 
Since the lattice is uniform, the local external potential (which would correspond to the nuclear potential in a conventional quantum chemical calculation) operator is proportional to the electron counting operator [see the last term on the right-hand side of Eq.~(\ref{eq:1D_Hubbard_Hamilt})], 
\be
\hat{N}=\sum^{L-1}_{i=0}\sum_{\sigma=\uparrow,\downarrow}\hat{n}_{i\sigma}.
\ee
The uniform value of the external potential can be rewritten as 
\be\label{eq:constant_ext_pot}
v_{\rm ext}=-\mu,
\ee
where the chemical potential $\mu$ controls the number of electrons $N$ or, equivalently, the uniform density $n=N/L$ in the lattice. In this case, $\hat{H}$ is actually a (zero-temperature) grand canonical Hamiltonian. For convenience, we rewrite the hopping operator as follows,
\be\label{eq:hopping_final_form_sumij}
\hat{T}\equiv\sum^{L-1}_{i,j=0}\sum_{\sigma=\uparrow,\downarrow}t_{ij}\hat{c}^\dagger_{i\sigma}\hat{c}_{j\sigma},
\ee
where
\be\label{eq:hopping_matrix}
t_{ij}=-t\left(\delta_{j(i+1)}+\delta_{i(j+1)}\right),
\ee
and $t_{(L-1)0}=t_{0(L-1)}=-t$. From now on the bounds in the summations over the full lattice will be dropped, for simplicity:
\be
\sum_i\equiv \sum^{L-1}_{i=0}.
\ee
Note that the quantum embedding strategies discussed in the present work can be extended to more general (quantum chemical, in particular) Hamiltonians~\cite{wouters2016practical}. For that purpose, the true {\it ab initio} Hamiltonian should be written in a localized molecular orbital basis, thus leading to the more general Hamiltonian expression,
\be
\hat{H}=\sum_{\sigma}\sum_{ij}h_{ij}\hat{c}^\dagger_{i\sigma}\hat{c}_{j\sigma}+\dfrac{1}{2}\sum_{\sigma,\sigma'}\sum_{ijkl}\langle ij\vert kl\rangle
\hat{c}^\dagger_{i\sigma}\hat{c}^\dagger_{j\sigma'}\hat{c}_{l\sigma'}\hat{c}_{k\sigma},
\ee
where $h_{ij}$ and $\langle ij\vert kl\rangle$ are the (kinetic and nuclear attraction) one-electron and two-electron repulsion integrals, respectively. Using a localized orbital basis allows for the decomposition of the molecule under study into fragments that can be embedded afterward~\cite{wouters2016practical}. In the following, we will work with the simpler Hamiltonian of Eq.~(\ref{eq:1D_Hubbard_Hamilt}), as a proof of concept.\\

\subsection{Review of Ht-DMFET}\label{subsec:review_Ht-DMFET}

For the sake of clarity and completeness, a review of Ht-DMFET~\cite{sekaran2021} is presented in the following subsections. Various ingredients (operators and reduced quantities) that will be used later on in Sec.~\ref{subsec:exact_dfe_dft} in the derivation of a formally exact density-functional embedding theory (which is the main outcome of this work) are introduced. Real algebra will be used. For simplicity, we focus on the embedding of a single impurity. A multiple-impurity extension of the theory can be obtained from a block Householder transformation~\cite{sekaran2021,AML99_Rotella_Block_Householder_transf}. Unlike in the exact reformulation of the theory which is proposed in the following Sec.~\ref{subsec:exact_dfe_dft} and where the chemical potential $\mu$ controls the density of the uniform lattice, the total number of electrons will be {\it fixed} to the value $N$ in the present section. In other words, the uniform density is set to $n=N/L$ and $\mu$ is an arbitrary constant (that could be set to zero).

\subsubsection{Exact non-interacting embedding}\label{subsubsec:non_int_embedding}

Let us first consider the particular case of a non-interacting ($U=0$) lattice for which Ht-DMFET is exact~\cite{sekaran2021}. As it will be applied later on (in Sec.~\ref{subsec:exact_dfe_dft}) to the auxiliary KS lattice, it is important to highlight the key features of the non-interacting embedding. Following Ref.~\cite{sekaran2021}, we label as $i=0$ one of the localized (lattice site in the present case) spin-orbital $\ket{\chi^\sigma_0}\equiv \hat{c}_{0\sigma}^\dagger\ket{\rm vac}$ [we denote $\ket{\rm vac}$ the vacuum state of second quantization] that, ultimately, will become the so-called {\it embedded impurity}. The ingredient that is central in Ht-DMFET is the (one-electron reduced) density matrix of the full system in the lattice representation, \ie,  
\be\label{eq:1RDM_lattice}
{\bm \gamma}^{\uparrow}={\bm \gamma}^{\downarrow}={\bm \gamma}\equiv\gamma_{ij}=\mel{\Phi}{\hat{c}_{i\sigma}^\dagger\hat{c}_{j\sigma}}{\Phi},
\ee
where we restrict ourselves to closed-shell singlet ground states $\ket{\Phi}$, for simplicity. 
Note that 
\be\label{eq:gamma_00_filling}
\gamma_{00}=\dfrac{n}{2}=\dfrac{N}{2L}
\ee
is the uniform lattice filling per spin.
Since the full lattice will always be described with a single Slater determinant in the following, the density matrix ${\bm \gamma}$ will always be {\it idempotent}. The latter is used to construct the Householder unitary transformation which, once it has been applied to the one-electron lattice space, defines the so-called {\it bath} spin-orbital with which the impurity will ultimately be exclusively entangled. More explicitly, the Householder transformation matrix \be\label{eq:P_from_HH_vec}
{\bm P}={\bm I}-2\bv\bv^{\dagger}\equiv P_{ij}=\delta_{ij}-2{\rm v}_i{\rm v}_j,
\ee
where ${\bm I}$ is the identity matrix, is a functional of the density matrix, \ie, 
\be\label{eq:P_functional_oneRDM}
{\bm P}\equiv{\bm P}\left[{\bm \gamma}\right],
\ee
where the density-matrix-functional Householder vector components read as~\cite{sekaran2021}
\be\label{eq:v0_zero}
{\rm v}_0&=&0,
\\
\label{eq:HH_vec_compt1}
{\rm v}_1&=&\dfrac{\gamma_{10}-\tilde{\gamma}_{10}}{\sqrt{2\tilde{\gamma}_{10}\left(\tilde{\gamma}_{10}-\gamma_{10}\right)}},
\\
\label{eq:HH_vec_compt_i_larger2}
{\rm v}_i&\underset{i\geq 2}{=}&\dfrac{\gamma_{i0}}{\sqrt{2\tilde{\gamma}_{10}\left(\tilde{\gamma}_{10}-\gamma_{10}\right)}},
\ee
with
\be\label{eq:gamma01_tilde}
\tilde{\gamma}_{10}=-{\rm sgn}\left(\gamma_{10}\right)\sqrt{\sum_{j>0}\gamma^2_{j0}},
\ee
and 
\be\label{eq:normalization_HH_vec}
{\bv}^\dagger\bv=\sum_{i\geq 1}{\rm v}^2_i=1.
\ee
Note that, in the extreme case of a two-site lattice, the denominator in Eqs.~(\ref{eq:HH_vec_compt1}) and (\ref{eq:HH_vec_compt_i_larger2}) is still well defined and it does not vanish. Indeed, by construction [see Eq.~(\ref{eq:gamma01_tilde})],
\be
\tilde{\gamma}_{10}&\underset{\tiny\left\{\gamma_{j0}\overset{j>1}{=}0\right\}}{=}&-{\rm sgn}\left(\gamma_{10}\right)\abs{\gamma_{10}}=-\gamma_{10}
\ee
in this case, thus leading to $\tilde{\gamma}_{10}\left(\tilde{\gamma}_{10}-\gamma_{10}\right)=2\gamma_{10}^2>0$.
Note also that ${\bm P}$ is hermitian and unitary, \ie, 
${\bm P}={\bm P}^\dagger$ and
\be\label{eq:unitary_transf}
{\bm P}^2={\bm P}{\bm P}^\dagger={\bm P}^\dagger{\bm P}={\bm I}.
\ee
The bath spin-orbital $\ket{\varphi^\sigma_{\rm bath}}$ is then constructed as follows in second quantization,
\be
\ket{\varphi^\sigma_{\rm bath}}:=\hat{d}_{1\sigma}^\dagger\ket{\rm vac},
\ee
where, according to Eqs.~(\ref{eq:P_from_HH_vec}) and (\ref{eq:v0_zero}),
\be\label{eq:bath_expansion}
\begin{split}
\hat{d}_{1\sigma}^\dagger&=\sum_{k}P_{1k}\hat{c}_{k\sigma}^\dagger
\\
&=\hat{c}_{1\sigma}^\dagger-2{\rm v}_1\sum_{k\geq 1}{\rm v}_k\hat{c}_{k\sigma}^\dagger.
\end{split}
\ee
More generally, the entire lattice space can be Householder-transformed as follows,
\be\label{eq:Householder_creation_ops}
\hat{d}_{i\sigma}^\dagger\underset{0\leq i\leq L-1}{=}\sum_k P_{ik}\hat{c}_{k\sigma}^\dagger,
\ee
and the back transformation simply reads as
\be\label{eq:from_HH_to_lattice_rep}
\sum_i P_{li}\hat{d}_{i\sigma}^\dagger=\sum_{ik}P_{li}P_{ik}\hat{c}_{k\sigma}^\dagger=\sum_{k}\left[{\bm P}^2\right]_{lk}\hat{c}_{k\sigma}^\dagger=\hat{c}_{l\sigma}^\dagger.
\ee
We stress that the impurity is invariant under the Householder transformation, \ie,
\be\label{eq:imp_invariant_underHH_transf}
\hat{d}_{0\sigma}^\dagger=\hat{c}_{0\sigma}^\dagger,
\ee
and, according to the Appendix, the Householder-transformed density matrix elements involving the impurity can be simplified as follows,
\be\label{eq:simplified_tilde_gamma0j}
\mel{\Phi}{\hat{d}_{j\sigma}^\dagger\hat{d}_{0\sigma}}{\Phi}=\gamma_{j0}-{\rm v}_j\sqrt{2\tilde{\gamma}_{10}\left(\tilde{\gamma}_{10}-\gamma_{10}\right)}.
\ee
As readily seen from Eqs.~(\ref{eq:HH_vec_compt1}) and (\ref{eq:simplified_tilde_gamma0j}), the matrix element $\tilde{\gamma}_{10}$ introduced in Eq.~(\ref{eq:gamma01_tilde}) is in fact the bath-impurity element of the density matrix in the Householder representation: 
\be
\mel{\Phi}{\hat{d}_{1\sigma}^\dagger\hat{d}_{0\sigma}}{\Phi}=\tilde{\gamma}_{10}.
\ee
If we denote 
\be\label{eq:1RDM_with_d_ops}
\tilde{\bm \gamma}\equiv \tilde{\gamma}_{ij}= \mel{\Phi}{\hat{d}_{i\sigma}^\dagger\hat{d}_{j\sigma}}{\Phi}=\sum_{kl}P_{ik}\gamma_{kl}P_{lj}\equiv {\bm P}{\bm \gamma}{\bm P} 
\ee
the full Householder-transformed density matrix, we do readily see from Eqs.~(\ref{eq:HH_vec_compt_i_larger2}) and (\ref{eq:simplified_tilde_gamma0j}) that the impurity is exclusively entangled with the bath, \ie, 
\be\label{eq:imp_disconnected_from_Henv}
\tilde{\gamma}_{i0}\underset{i\geq 2}{=}0,
\ee
by construction~\cite{sekaran2021}. As $\tilde{\bm \gamma}$ inherits the idempotency of ${\bm \gamma}$ through the unitary Householder transformation, we deduce from Eq.~(\ref{eq:imp_disconnected_from_Henv}) that
\be
\tilde{\gamma}_{i0}=\left[\tilde{\bm \gamma}^2\right]_{i0}=\sum_j\tilde{\gamma}_{ij}\tilde{\gamma}_{j0}=\tilde{\gamma}_{i0}\tilde{\gamma}_{00}+\tilde{\gamma}_{i1}\tilde{\gamma}_{10},
\ee
or, equivalently,
\be
\tilde{\gamma}_{i1}=\dfrac{\tilde{\gamma}_{i0}\left(1-\tilde{\gamma}_{00}\right)}{\tilde{\gamma}_{10}},
\ee
thus leading to [see Eq.~(\ref{eq:imp_disconnected_from_Henv})]
\be\label{eq:bath_entangled_with_imp}
\tilde{\gamma}_{i1}\underset{i\geq 2}{=}0,
\ee
and 
\be\label{eq:2e_in_cluster}
\tilde{\gamma}_{00}+\tilde{\gamma}_{11}=1.
\ee
Eqs.~(\ref{eq:bath_entangled_with_imp}) and (\ref{eq:2e_in_cluster}) simply indicate that, by construction~\cite{sekaran2021}, the bath is itself entangled exclusively with the impurity, and the Householder ``impurity+bath'' cluster, which is disconnected from its environment, contains exactly two electrons (one per spin). Therefore, the Householder cluster sector of the density matrix can be described exactly by a {\it two-electron} Slater determinant $\Phi^{\mathcal{C}}$:
\be\label{eq:cluster_sector_Psi_representable}
\tilde{\gamma}_{ij}\underset{0\leq i,j\leq 1}{=}\mel{\Phi^{\mathcal{C}}}{\hat{d}_{i\sigma}^\dagger\hat{d}_{j\sigma}}{\Phi^{\mathcal{C}}}.
\ee
Note that, in the Householder representation, the lattice ground-state determinant reads as $\Phi\equiv \Phi^{\mathcal{C}}\Phi_{\rm core}$, where the cluster's determinant $\Phi^{\mathcal{C}}$ is disentangled from the core one $\Phi_{\rm core}$. Once the cluster's block of the density matrix has been diagonalized, we obtain the sole occupied orbital that overlaps with the impurity, exactly like in DMET~\cite{wouters2016practical}. In other words, for non-interacting (or mean-field-like descriptions of) electrons, the Ht-DMFET construction of the bath is equivalent (although simpler) to that of DMET. We refer the reader to Ref.~\cite{sekaran2021} for a more detailed comparison of the two approaches.\\ 

\subsubsection{Non-interacting embedding Hamiltonian}

As the Householder cluster is strictly disconnected from its environment in the non-interacting case, it is exactly described by the two-electron ground state $\ket{\Phi^{\mathcal{C}}}$ of the Householder-transformed hopping operator (that we refer to as kinetic energy operator from now on, like in  DFT for lattices~\cite{DFT_ModelHamiltonians,senjean2018site}) on projected onto the cluster~\cite{sekaran2021}, \ie, 
\be\label{eq:non-int_cluster_SE}
\hat{\mathcal{T}}^{\mathcal{C}}\ket{\Phi^{\mathcal{C}}}=\mathcal{E}_{\rm s}^{\mathcal{C}}\ket{\Phi^{\mathcal{C}}},
\ee
where, according to Eqs.~(\ref{eq:hopping_final_form_sumij}) and (\ref{eq:from_HH_to_lattice_rep}),
\be
\hat{\mathcal{T}}^{\mathcal{C}}=\sum_{ij}\sum_{\sigma=\uparrow,\downarrow}t_{ij}
\sum^1_{k,l=0}P_{ik}P_{jl}\hat{d}_{k\sigma}^\dagger\hat{d}_{l\sigma}.
\ee
For convenience, we will separate in $\hat{\mathcal{T}}^{\mathcal{C}}$ the physical per-site kinetic energy operator [see Eq.~(\ref{eq:hopping_operator})],
\be
\hat{t}_{01}=-t\sum_{\sigma=\uparrow,\downarrow}\left(\hat{c}^\dagger_{0\sigma}\hat{c}_{1\sigma}
+\hat{c}_{1\sigma}^\dagger\hat{c}_{0\sigma}
\right),
\ee
from the correction induced (within the cluster) by the Householder transformation:
\be\label{eq:hc_t01_plus_corr}
\hat{\tau}^{\mathcal{C}}=
\hat{\mathcal{T}}^{\mathcal{C}}-\hat{t}_{01}.
\ee
Note that, since $t_{00}=0$, $\hat{\tau}^{\mathcal{C}}$ can be expressed more explicitly as follows,
\be
\begin{split}
\hat{\tau}^{\mathcal{C}}
&=\sum_{\sigma=\uparrow,\downarrow}\left(\sum_{ij}P_{i1}P_{j0}t_{ij}\right)\left[\hat{d}_{0\sigma}^\dagger\hat{d}_{1\sigma}+\hat{d}_{1\sigma}^\dagger\hat{d}_{0\sigma}\right]\\
&\quad +\sum_{\sigma=\uparrow,\downarrow}\left(\sum_{ij}P_{i1}P_{j1}t_{ij}\right)\hat{d}_{1\sigma}^\dagger\hat{d}_{1\sigma}-\hat{t}_{01}
\\
&=\sum_{\sigma=\uparrow,\downarrow}\left(\sum_iP_{i1}t_{i0}\right)\left[\hat{c}_{0\sigma}^\dagger\hat{d}_{1\sigma}+\hat{d}_{1\sigma}^\dagger\hat{c}_{0\sigma}\right]
\\
&\quad +\sum_{\sigma=\uparrow,\downarrow}\left(\sum_{ij}P_{i1}P_{j1}t_{ij}\right)\hat{d}_{1\sigma}^\dagger\hat{d}_{1\sigma}-\hat{t}_{01}
\\
&=\sum_{\sigma=\uparrow,\downarrow}t_{10}\left[\hat{c}_{0\sigma}^\dagger\hat{d}_{1\sigma}+\hat{d}_{1\sigma}^\dagger\hat{c}_{0\sigma}\right]
\\
&\quad -2{\rm v}_1\sum_{\sigma=\uparrow,\downarrow}\left(\sum_i{\rm v}_it_{i0}\right)\left[\hat{c}_{0\sigma}^\dagger\hat{d}_{1\sigma}+\hat{d}_{1\sigma}^\dagger\hat{c}_{0\sigma}\right]
\\
&\quad +\sum_{\sigma=\uparrow,\downarrow}\left(\sum_{ij}P_{i1}P_{j1}t_{ij}\right)\hat{d}_{1\sigma}^\dagger\hat{d}_{1\sigma}-\hat{t}_{01},
\end{split}
\ee
thus leading to
\be\label{eq:simplified_tauC}
\begin{split}
\hat{\tau}^{\mathcal{C}}
&=2t{\rm v}_1\sum_{\sigma=\uparrow,\downarrow}\sum_{k\geq 1}{\rm v}_k\left[\hat{c}_{0\sigma}^\dagger\hat{c}_{k\sigma}+\hat{c}_{k\sigma}^\dagger\hat{c}_{0\sigma}\right]
\\
&\quad -2{\rm v}_1\sum_{\sigma=\uparrow,\downarrow}\left(\sum_i{\rm v}_it_{i0}\right)\left[\hat{c}_{0\sigma}^\dagger\hat{d}_{1\sigma}+\hat{d}_{1\sigma}^\dagger\hat{c}_{0\sigma}\right]
\\
&\quad+4\left(\sum_{ij}{\rm v}_i{\rm v}_j\left({\rm v}_1^2-\delta_{j1}\right)t_{ij}\right)\sum_{\sigma=\uparrow,\downarrow}\hat{d}_{1\sigma}^\dagger\hat{d}_{1\sigma},
\end{split}
\ee
where we used Eqs.~(\ref{eq:P_from_HH_vec}) and (\ref{eq:bath_expansion}), as well as the fact that $t_{11}=0$ and $t_{10}=-t$.
Note that, when no Householder transformation is performed (\ie, when ${\rm v}_i=0$ for $0\leq i\leq L-1$), the bath site simply corresponds to the nearest neighbor ($i=1$) of the impurity in the lattice [see Eq.~(\ref{eq:bath_expansion})] and, as readily seen from Eqs.~(\ref{eq:hc_t01_plus_corr}) and (\ref{eq:simplified_tauC}), the non-interacting cluster's Hamiltonian $\hat{\mathcal{T}}^{\mathcal{C}}$ reduces to $\hat{t}_{01}$.\\ 

Unlike in the interacting case, which is discussed in Sec.~\ref{sec:approx_int_embedding}, it is unnecessary to introduce an additional potential on the embedded impurity in order to ensure that it reproduces the correct lattice filling. Indeed, according to Eqs.~(\ref{eq:gamma_00_filling}), (\ref{eq:v0_zero}), (\ref{eq:simplified_tilde_gamma0j}), (\ref{eq:1RDM_with_d_ops}), \ref{eq:cluster_sector_Psi_representable}),
\be\label{eq:occ_embedded_imp_lattice_equal}
\mel{\Phi^{\mathcal{C}}}{\hat{c}_{0\sigma}^\dagger\hat{c}_{0\sigma}}{\Phi^{\mathcal{C}}}=\mel{\Phi^{\mathcal{C}}}{\hat{d}_{0\sigma}^\dagger\hat{d}_{0\sigma}}{\Phi^{\mathcal{C}}}=n/2.
\ee
This constraint is automatically fulfilled when Householder transforming the kinetic energy operator $\hat{T}$ of the full lattice, thanks to the local potential contribution on the bath [see the last term on the right-hand side of Eq.~(\ref{eq:simplified_tauC})].
Interestingly, the true (non-interacting in this case) per-site energy of the lattice can be determined solely from $\Phi^{\mathcal{C}}$. Indeed, according to Eq.~(\ref{eq:1RDM_lattice}), the per-site kinetic energy can be evaluated from the lattice ground-state wave function $\Phi$ as follows, 
\be\label{eq:true_per-site_ener_lattice}
\mel{\Phi}{\hat{t}_{01}}{\Phi}=-4t\gamma_{10}.
\ee
When rewritten in the Householder representation, Eq.~(\ref{eq:true_per-site_ener_lattice}) gives [see Eqs.~(\ref{eq:from_HH_to_lattice_rep}), (\ref{eq:imp_disconnected_from_Henv}), and (\ref{eq:cluster_sector_Psi_representable})]
\be\label{eq:ener_from_lattice_to_cluster}
\begin{split}
\mel{\Phi}{\hat{t}_{01}}{\Phi}
&=-4t\sum_iP_{1i}\tilde{\gamma}_{i0}
\\
&=-4t\sum_{0\leq i\leq 1}P_{1i}\tilde{\gamma}_{i0}
\\
&=-4t\sum_{0\leq i\leq 1}P_{1i}\mel{\Phi^{\mathcal{C}}}{\hat{d}_{i\sigma}^\dagger\hat{d}_{0\sigma}}{\Phi^{\mathcal{C}}}
\\
&=-4t\sum_{i}P_{1i}\mel{\Phi^{\mathcal{C}}}{\hat{d}_{i\sigma}^\dagger\hat{c}_{0\sigma}}{\Phi^{\mathcal{C}}},
\end{split}
\ee
where we used Eq.~(\ref{eq:imp_invariant_underHH_transf}) and the fact that $\hat{d}_{i\sigma}\ket{\Phi^{\mathcal{C}}}\overset{i>1}{=}0$, since $\Phi^{\mathcal{C}}$ is constructed within the cluster. We finally recover from Eq.~(\ref{eq:ener_from_lattice_to_cluster}) the following equality~\cite{sekaran2021},
\be\label{eq:per_site_kin_ener_from_cluster}
\begin{split}
\mel{\Phi}{\hat{t}_{01}}{\Phi}&=-4t\mel{\Phi^{\mathcal{C}}}{\hat{c}_{1\sigma}^\dagger\hat{c}_{0\sigma}}{\Phi^{\mathcal{C}}}
\\
&=\mel{\Phi^{\mathcal{C}}}{\hat{t}_{01}}{\Phi^{\mathcal{C}}},
\end{split}\ee
which drastically (and exactly) simplifies the evaluation of non-interacting energies for lattices.

\subsubsection{Approximate interacting embedding}\label{sec:approx_int_embedding}

The simplest (approximate) extension of Ht-DMFET to interacting electrons consists in introducing the on-impurity-site two-electron repulsion operator $\hat{U}_0$ into the non-interacting Householder cluster's Hamiltonian of Eq.~(\ref{eq:non-int_cluster_SE}), by analogy with DMET~\cite{knizia2012density,sekaran2021}. In such a (standard) scheme, the interaction is treated {\it on top} of the non-interacting embedding. Unlike in the non-interacting case, it is necessary to introduce a chemical potential $\tilde{\mu}^{\rm imp}$ on the embedded impurity in order to ensure that it reproduces the correct lattice filling $N/L$~\cite{sekaran2021}, \ie, 
\be
\expval{\hat{n}_0}_{\Psi^{\mathcal{C}}}=N/L,
\ee
where the two-electron cluster's ground-state wave function $\Psi^{\mathcal{C}}$ fulfills the following interacting Schr\"{o}dinger equation:
\be\label{eq:int_cluster_SE}
\left(\hat{\mathcal{T}}^{\mathcal{C}}+\hat{U}_0-\tilde{\mu}^{\rm imp}\hat{n}_0\right)\ket{\Psi^{\mathcal{C}}}=\mathcal{E}^{\mathcal{C}}\ket{\Psi^{\mathcal{C}}}.
\ee
The physical per-site energy (from which we remove the chemical potential contribution) is then evaluated as follows:  
\be\label{eq:per_site_ener_HtDMFET}
\left(E+\mu N\right)/{L}\underset{\rm{Ht-DMFET}}{\approx}\mel{\Psi^{\mathcal{C}}}{\hat{t}_{01}+\hat{U}_0}{\Psi^{\mathcal{C}}}.
\ee
Let us stress that, in Ht-DMFET, the cluster is designed from a single determinantal (non-interacting in the present case) lattice wave function, like in regular DMET calculations~\cite{wouters2016practical}. In other words, the Householder transformation is constructed from an idempotent density matrix. Moreover, the interacting cluster is described as a {\it closed} (two-electron) subsystem. As shown for small Hubbard rings, the exact interacting cluster is in principle an open subsystem~\cite{sekaran2021}. It rigorously contains two electrons only at half filling, as a consequence of the hole-particle symmetry of the Hubbard lattice Hamiltonian~\cite{sekaran2021}.\\ 

Note finally that, if we Householder transform the two-electron repulsion operator $\hat{U}$ of the full lattice, one can in principle take into account its complete projection onto the cluster. It means that the interaction on the bath site could be added to the Hamiltonian in Eq.~(\ref{eq:int_cluster_SE}). For simplicity, we will focus in the following on the (so-called) non-interacting bath formulation of the theory, which is described by Eq.~(\ref{eq:int_cluster_SE}). Let us finally mention that, in the present single-impurity embedding, DMET, DET, and Ht-DMFET are equivalent~\cite{sekaran2021}.

\subsection{Exact density-functional embedding}\label{subsec:exact_dfe_dft}

We will show in the following that, once it has been merged with KS-DFT, Ht-DMFET can be made formally exact. For clarity, we start with reviewing briefly KS-DFT for lattice Hamiltonians in Sec.~\ref{subsubsec:KS-DFT_lattices}. A multi-determinantal extension of the theory based on the interacting Householder cluster's wave function is then proposed in Sec.~\ref{subsubsec:exact_DFE}.    

\subsubsection{KS-DFT for uniform lattices}\label{subsubsec:KS-DFT_lattices}

According to the Hohenberg--Kohn (HK) variational principle~\cite{hktheo}, which is applied in this work to lattice Hamiltonians~\cite{DFT_ModelHamiltonians}, the ground-state energy of the full lattice can be determined as follows,
\be\label{eq:HK_var_principle_full_lattice}
E=\min_n\left\{F(n)+v_{\rm ext}nL\right\},
\ee
where the HK density functional reads as
\be
F(n)=\mel{\Psi(n)}{\hat{T}+\hat{U}}{\Psi(n)},
\ee
and $\ket{\Psi(n)}$ is the lattice ground state with uniform density profile $n\overset{0\leq i< L}{=}\mel{\Psi(n)}{\hat{n}_i}{\Psi(n)}$. Strictly speaking, $F(n)$ is a function of the site occupation $n$, hence the name {\it site occupation functional theory} often given to DFT for lattices~\cite{DFT_ModelHamiltonians,senjean2018site}. Note that the ground-state energy $E$ is in fact a (zero-temperature) grand canonical energy since a change in uniform density $n$ induces a change in the number $N=nL$ of electrons. In the thermodynamic $N\rightarrow+\infty$ and $L\rightarrow+\infty$ limit, with $N/L$ fixed to $n$, one can in principle describe {\it continuous} variations in $n$ with a pure-state wave function ${\Psi(n)}$. The derivations that follow will be based on this assumption. If we introduce the per-site analog of the HK functional,
\be\label{eq:per_site_HK_func}
f(n)=F(n)/L=\mel{\Psi(n)}{\hat{t}_{01}+\hat{U}_0}{\Psi(n)},
\ee
and use the notation of Eq.~(\ref{eq:constant_ext_pot}), then Eq.~(\ref{eq:HK_var_principle_full_lattice}) becomes 
\be
E/L\equiv E(\mu)/L=\min_n\left\{f(n)-\mu n\right\},
\ee
and the minimizing density $n(\mu)$ fulfills the following stationarity condition:
\be\label{eq:true_chem_pot_from_DFT}
\mu=\left.\dfrac{\partial f(n)}{\partial n}\right|_{n=n(\mu)}.
\ee
In the conventional KS formulation of DFT, the per-site HK functional is decomposed as follows,
\be\label{eq:KS_decomp}
f(n)=t_{\rm s}(n)+e_{\rm Hxc}(n),
\ee
where 
\be\label{eq:per_site_ts}
t_{\rm s}(n)=\mel{\Phi(n)}{\hat{t}_{01}}{\Phi(n)}=\dfrac{1}{L}\mel{\Phi(n)}{\hat{T}}{\Phi(n)}
\ee
is the (per-site) analog for lattices of the non-interacting kinetic energy functional, and the Hxc density functional reads as~\cite{DFT_ModelHamiltonians}
\be
e_{\rm Hxc}(n)=\dfrac{U}{4}n^2+e_{\rm c}(n),
\ee
where $e_{\rm c}(n)$ is the exact (per-site) correlation energy functional of the interacting lattice. The (normalized) density-functional lattice KS determinant ${\Phi(n)}$ 
fulfills the (non-interacting) KS equation
\be
\left(\hat{T}-\mu_{\rm s}(n)\hat{N}\right)\ket{\Phi(n)}=\mathcal{E}_{\rm s}(n)\ket{\Phi(n)},
\ee
so that [see Eq.~(\ref{eq:per_site_ts})] 
\be
\begin{split}
\dfrac{\partial t_{\rm s}(n)}{\partial n}&=\dfrac{2}{L}\mel{\frac{\partial \Phi(n)}{\partial n}}{\hat{T}}{\Phi(n)}
\\
&=\dfrac{2\mu_{\rm s}(n)}{L}\mel{\frac{\partial \Phi(n)}{\partial n}}{\hat{N}}{\Phi(n)}
\\
&=\dfrac{\mu_{\rm s}(n)}{L}\dfrac{\partial (nL)}{\partial n}
\\
&=\mu_{\rm s}(n),
\end{split}
\ee
since $\mel{\Phi(n)}{\hat{N}}{\Phi(n)}=N=nL$.
Thus we recover from Eqs.~(\ref{eq:true_chem_pot_from_DFT}) and (\ref{eq:KS_decomp}) the well-known relation between the physical and KS chemical potentials:
\be\label{eq:KS_pot_decomp}
\mu_{\rm s}(n(\mu))\equiv \mu_{\rm s}=\mu-v_{\rm Hxc}, 
\ee
where the density-functional Hxc potential reads as $v_{\rm Hxc}=v_{\rm Hxc}(n(\mu))$ with
\be\label{eq:Hxc_pot_from_eHxc}
v_{\rm Hxc}(n)=
\dfrac{\partial e_{\rm Hxc}(n)}{\partial n}.
\ee
Note that the exact non-interacting density-functional chemical potential can be expressed analytically as follows~\cite{lima2003density}:
\be\label{eq:KS_dens_func_chemical_potential}
\mu_{\rm s}(n) = -2t\cos\left(\frac{\pi}{2}n\right).
\ee
Capelle and coworkers~\cite{lima2003density,DFT_ModelHamiltonians} have designed a local density approximation (LDA) to $e_{\rm Hxc}(n)$ on the basis of exact Bethe Ansatz (BA) solutions~\cite{lieb_absence_1968} (the functional is usually referred to as BALDA).\\

Unlike in conventional {\it ab initio} DFT, the Hxc functional of lattice Hamiltonians is not truly universal in the sense that it is universal for a given choice of (hopping) one-electron and two-electron repulsion operators. In other words, the Hxc functional does not depend on the (possibly non-uniform) one-electron local potential operator $\sum_i v_{{\rm ext},i}\hat{n}_i$, which is the analog for lattices of the nuclear potential in molecules, but it is $t$- and $U$-dependent and, in the present case, it should be designed specifically for the 1D Hubbard model. Even though BALDA can be extended to higher dimensions~\cite{Vilela_2019}, there is no general strategy for constructing (localized) orbital-occupation functional approximations, thus preventing direct applications to quantum chemistry~\cite{fromager2015exact}, for example. Turning ultimately to a potential-functional theory, as proposed in Sec.~\ref{sec:LPFET}, is appealing in this respect. With this change of paradigm, which is the second key result of the paper, the Hxc energy and potential become implicit functionals of the density, and they can be evaluated from a (few-electron) correlated wave function through a quantum embedding procedure.     

\subsubsection{Density-functional interacting cluster}\label{subsubsec:exact_DFE}

We propose in this section an alternative formulation of DFT based on the interacting Householder cluster introduced in Sec.~\ref{sec:approx_int_embedding}. For that purpose, we consider the following {\it exact} decomposition,
\be\label{eq:decomp_f_from_cluster}
f(n)=f^{\mathcal{C}}(n)+\overline{e}_{\rm c}(n),
\ee
where the Householder cluster HK functional
\be\label{eq:f_cluster_func}
f^{\mathcal{C}}(n)=\mel{\Psi^{\mathcal{C}}(n)}{\hat{t}_{01}+\hat{U}_0}{\Psi^{\mathcal{C}}(n)}
\ee
is evaluated from the two-electron cluster density-functional wave function ${\Psi^{\mathcal{C}}(n)}$, and $\overline{e}_{\rm c}(n)$ is the complementary correlation density functional that describes the correlation effects of the Householder cluster's environment on the embedded impurity. Note that, according to Sec.~\ref{sec:approx_int_embedding}, $\ket{ \Psi^{\mathcal{C}}(n) }$ fulfills the following Schr\"{o}dinger-like equation,
\be\label{eq:dens_func_int_cluster_SE}
\hat{\mathcal{H}}^{\mathcal{C}}(n)\ket{\Psi^{\mathcal{C}}(n)}=\mathcal{E}^{\mathcal{C}}(n)\ket{\Psi^{\mathcal{C}}(n)},
\ee
where (we use the same notations as in Sec.~\ref{sec:approx_int_embedding})
\be\label{eq:dens_func_cluster_hamilt}
\hat{\mathcal{H}}^{\mathcal{C}}(n)\equiv \hat{\mathcal{T}}^{\mathcal{C}}(n)+\hat{U}_0-\tilde{\mu}^{\rm imp}(n)\,\hat{n}_0
\ee
and 
\be\label{eq:Tcluster_op_dens_func}
\hat{\mathcal{T}}^{\mathcal{C}}(n)\equiv\hat{t}_{01}+\hat{\tau}^{\mathcal{C}}(n).
\ee
The dependence in $n$ of the (projected-onto-the-cluster) Householder-transformed kinetic energy operator $\hat{\mathcal{T}}^{\mathcal{C}}(n)$ comes from the fact that the KS lattice density matrix ${\bm \gamma}(n)\equiv\mel{\Phi(n)}{\hat{c}_{i\sigma}^\dagger\hat{c}_{j\sigma}}{\Phi(n)}$ (on which the Householder transformation is based) is, like the KS determinant $\Phi(n)\equiv \Phi^{\mathcal{C}}(n)\Phi_{\rm core}(n)$ of the lattice, a functional of the uniform density $n$. On the other hand, for a given uniform lattice density $n$, the local potential $-\tilde{\mu}^{\rm imp}(n)$ is adjusted on the embedded impurity  such that the interacting cluster reproduces $n$, \ie, 
\be\label{eq:dens_constraint_int_cluster}
\mel{\Psi^{\mathcal{C}}(n)}{\hat{n}_0}{\Psi^{\mathcal{C}}(n)}=n.
\ee

Interestingly, on the basis of the two decompositions in Eqs.~(\ref{eq:KS_decomp}) and (\ref{eq:decomp_f_from_cluster}), and Eq.~(\ref{eq:f_cluster_func}), we can relate the exact Hxc functional to the density-functional Householder cluster as follows,
\be
e_{\rm Hxc}(n)&=\mel{\Psi^{\mathcal{C}}(n)}{\hat{t}_{01}+\hat{U}_0}{\Psi^{\mathcal{C}}(n)}
-t_{\rm s}(n)+\overline{e}_{\rm c}(n),
\ee
where, as shown in Eq.~(\ref{eq:per_site_kin_ener_from_cluster}), the per-site non-interacting kinetic energy can be determined exactly from the two-electron cluster's part $\Phi^{\mathcal{C}}(n)$ of the KS lattice determinant $\Phi(n)$, \ie,
\be
t_{\rm s}(n)=\mel{\Phi^{\mathcal{C}}(n)}{\hat{t}_{01}}{\Phi^{\mathcal{C}}(n)},
\ee
thus leading to the final expression
\be\label{eq:final_exp_eHxc_from_cluster}
\begin{split}
e_{\rm Hxc}(n)&=\mel{\Psi^{\mathcal{C}}(n)}{\hat{t}_{01}+\hat{U}_0}{\Psi^{\mathcal{C}}(n)}
-\mel{\Phi^{\mathcal{C}}(n)}{\hat{t}_{01}}{\Phi^{\mathcal{C}}(n)}+\overline{e}_{\rm c}(n).
\end{split}
\ee
Note that, according to Eqs.~(\ref{eq:non-int_cluster_SE}) and (\ref{eq:hc_t01_plus_corr}), $\Phi^{\mathcal{C}}(n)$ fulfills the KS-like equation
\be\label{eq:dens_func_cluster_KS_eq}
\left(\hat{t}_{01}+\hat{\tau}^{\mathcal{C}}(n)\right)\ket{\Phi^{\mathcal{C}}(n)}=\mathcal{E}_{\rm s}^{\mathcal{C}}(n)\ket{\Phi^{\mathcal{C}}(n)},
\ee
where the Householder transformation ensures that $\mel{\Phi^{\mathcal{C}}(n)}{\hat{n}_0}{\Phi^{\mathcal{C}}(n)}=n$ [see Eq.~(\ref{eq:occ_embedded_imp_lattice_equal})].\\

We will now establish a clearer connection between the KS lattice system and the Householder cluster {\it via} the evaluation of the Hxc density-functional potential in the lattice. According to Eqs.~(\ref{eq:Hxc_pot_from_eHxc}) and (\ref{eq:final_exp_eHxc_from_cluster}), the latter can be expressed as follows,
\be
\begin{split}
v_{\rm Hxc}(n)&=2\mel{\frac{\partial\Psi^{\mathcal{C}}(n)}{\partial n}}{\hat{t}_{01}+\hat{U}_0}{\Psi^{\mathcal{C}}(n)}
\\
&\quad -2\mel{\frac{\partial\Phi^{\mathcal{C}}(n)}{\partial n}}{\hat{t}_{01}}{\Phi^{\mathcal{C}}(n)}+\frac{\partial \overline{e}_{\rm c}(n)}{\partial n},
\end{split}
\ee
or, equivalently [see Eqs.~(\ref{eq:dens_func_int_cluster_SE}), (\ref{eq:dens_constraint_int_cluster}), and (\ref{eq:dens_func_cluster_KS_eq})], 
\be\label{eq:Hxc_dens_pot_final}
\begin{split}
v_{\rm Hxc}(n)&=
\tilde{\mu}^{\rm imp}(n)
-2\mel{\frac{\partial\Psi^{\mathcal{C}}(n)}{\partial n}}{\hat{\tau}^{\mathcal{C}}(n)}{\Psi^{\mathcal{C}}(n)}
\\
&\quad +2\mel{\frac{\partial\Phi^{\mathcal{C}}(n)}{\partial n}}{\hat{\tau}^{\mathcal{C}}(n)}{\Phi^{\mathcal{C}}(n)}+\frac{\partial \overline{e}_{\rm c}(n)}{\partial n}.
\end{split}
\ee
If we introduce the following bi-functional of the density,
\be\label{eq:kinetic_corr_bi_func}
\begin{split}
\tau^{\mathcal{C}}_{\rm c}(n,\nu)&=\mel{\Psi^{\mathcal{C}}(\nu)}{\hat{\tau}^{\mathcal{C}}(n)}{\Psi^{\mathcal{C}}(\nu)}
-\mel{\Phi^{\mathcal{C}}(\nu)}{\hat{\tau}^{\mathcal{C}}(n)}{\Phi^{\mathcal{C}}(\nu)},
\end{split}
\ee
which can be interpreted as a kinetic correlation energy induced within the density-functional cluster by the Householder transformation and the interaction on the impurity, we obtain the final {\it exact} expression
\be\label{eq:final_vHxc_exp_from_muimp}
v_{\rm Hxc}(n)=
\tilde{\mu}^{\rm imp}(n)-\left.\frac{\partial \tau^{\mathcal{C}}_{\rm c}(n,\nu)}{\partial \nu}\right|_{\nu=n}+\frac{\partial \overline{e}_{\rm c}(n)}{\partial n},
\ee
which is the first key result of this paper.\\

Before turning Eq.~(\ref{eq:final_vHxc_exp_from_muimp}) into a practical self-consistent embedding method (see Sec.~\ref{sec:LPFET}), let us briefly discuss its physical meaning and connection with Ht-DMFET. As pointed out in Sec.~\ref{subsubsec:non_int_embedding}, the (density-functional) operator $\hat{\tau}^{\mathcal{C}}(n)$ is an auxiliary correction to the true per-site kinetic energy operator $\hat{t}_{01}$ which originates from the Householder-transformation-based embedding of the impurity. It is not physical and its impact on the impurity chemical potential $\tilde{\mu}^{\rm imp}(n)$, which is determined in the presence of $\hat{\tau}^{\mathcal{C}}(n)$ in the cluster's Hamiltonian [see Eqs.~(\ref{eq:dens_func_int_cluster_SE})-(\ref{eq:Tcluster_op_dens_func})], should be removed when evaluating the Hxc potential of the true lattice, hence the minus sign in front of the second term on the right-hand side of Eq.~(\ref{eq:final_vHxc_exp_from_muimp}). Finally, the complementary correlation potential $\partial \overline{e}_{\rm c}(n)/\partial n$ is in charge of recovering the electron correlation effects that were lost when considering an interacting cluster that is disconnected from its environment~\cite{sekaran2021}. We should stress at this point that, in Ht-DMFET (which is equivalent to DMET or DET when a single impurity is embedded~\cite{sekaran2021}), the following density-functional approximation is made:
\be\label{eq:DFA_in_HtDMFET}
\overline{e}_{\rm c}(n)\underset{\rm Ht-DMFET}{\approx}0,
\ee
so that the physical density-functional chemical potential is evaluated as follows~\cite{sekaran2021},
\be\label{eq:true_chem_pot_Ht-DMFET}
\mu(n)\underset{\rm Ht-DMFET}{\approx}\frac{\partial f^{\mathcal{C}}(n)}{\partial n}.
\ee
Interestingly, even though it is never computed explicitly in this context, the corresponding (approximate) Hxc potential simply reads as
\be
v_{\rm Hxc}(n)\underset{\rm{Ht-DMFET}}{\approx}\frac{\partial (f^{\mathcal{C}}(n)-t_{\rm s}(n))}{\partial n},
\ee
or, equivalently [see Eqs.~(\ref{eq:final_vHxc_exp_from_muimp}) and (\ref{eq:DFA_in_HtDMFET})],
\be\label{eq:approx_Hxc_pot_HtDMFET}
v_{\rm Hxc}(n)\underset{\rm{Ht-DMFET}}{\approx}
\tilde{\mu}^{\rm imp}(n)-\left.\frac{\partial \tau^{\mathcal{C}}_{\rm c}(n,\nu)}{\partial \nu}\right|_{\nu=n}.
\ee
Therefore, Ht-DMFET can be seen as an approximate formulation of KS-DFT where the Hxc potential is determined solely from the density-functional Householder cluster.  

\subsection{Local potential functional embedding theory}\label{sec:LPFET}

\begin{figure}[!htbp]
\begin{center}
\includegraphics[scale=0.4]{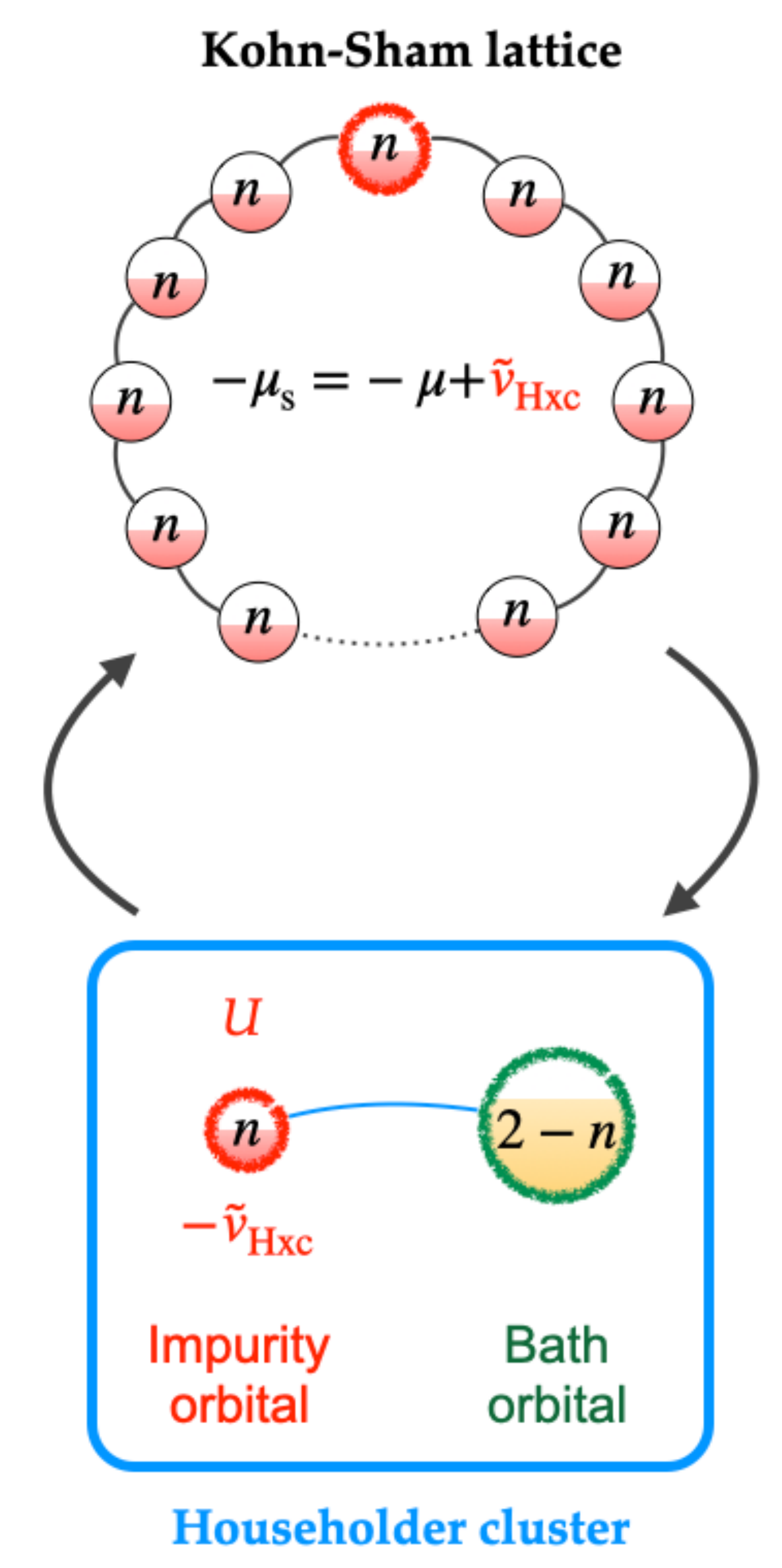}
\caption{Graphical representation of the LPFET procedure. Note that the {\it same} Hxc potential $\tilde{v}_{\rm Hxc}$ is used in the KS lattice and the embedding Householder cluster. It is optimized self-consistently in order to fulfill the density constraint of Eq.~(\ref{eq:sc_LPFET_eq}). See text for further details.}
\label{Fig:self-consistent-loop-scheme}
\end{center}
\end{figure}

 Until now the Householder transformation has been described as a functional of the uniform density $n$ or, more precisely, as a functional of the KS density matrix, which is itself a functional of the density. If we opt for a potential-functional reformulation of the theory, as suggested in the following, the Householder transformation becomes a functional of the KS chemical potential $\mu_{\rm s}$ instead, and, consequently, the Householder correction to the per-site kinetic energy operator within the cluster [see Eq.~(\ref{eq:Tcluster_op_dens_func})] is also a functional of $\mu_{\rm s}$:
 \be
 \hat{\tau}^{\mathcal{C}}(n)\rightarrow \hat{\tau}^{\mathcal{C}}(\mu_{\rm s}).
 \ee
 Similarly, the interacting cluster's wave function becomes a bi-functional of the KS {\it and} interacting embedded impurity chemical potentials:
 \be\label{eq:bifunctional_cluster_wfn}
 \Psi^{\mathcal{C}}(n)\rightarrow \Psi^{\mathcal{C}}\left(\mu_{\rm s},\tilde{\mu}^{\rm imp}\right).
 \ee
 In the exact theory, for a given chemical potential value $\mu$ in the true interacting lattice, both the KS lattice and the embedded impurity reproduce the interacting lattice density $n(\mu)$, \ie,
\be\label{eq:exact_dens_mapping_KS_cluster}
n(\mu)=n_{\rm lattice}^{\rm KS}\left(\mu-v_{\rm Hxc}\right)=n^{\mathcal{C}}\left(\mu-v_{\rm Hxc},\tilde{\mu}^{\rm imp}\right),
\ee
where
\be
n_{\rm lattice}^{\rm KS}(\mu_{\rm s})\equiv\expval{\hat{n}_0}_{\hat{T}-\mu_{\rm s}\hat{N}},
\ee
and
\be
\begin{split}
n^{\mathcal{C}}\left(\mu_{\rm s},\tilde{\mu}^{\rm imp}\right)
&=\expval{\hat{n}_0}_{\Psi^{\mathcal{C}}\left(\mu_{\rm s},\tilde{\mu}^{\rm imp}\right)}
\\
&\equiv
\expval{\hat{n}_0}_{\hat{t}_{01}+\hat{\tau}^{\mathcal{C}}(\mu_{\rm s})+\hat{U}_0-\tilde{\mu}^{\rm imp}\hat{n}_0},
\end{split}
\ee
with, according to Eq.~(\ref{eq:final_vHxc_exp_from_muimp}),
\be\label{eq:exact_muimp_expression}
\begin{split}
\tilde{\mu}^{\rm imp}&=\tilde{\mu}^{\rm imp}(n(\mu))
\\
&=v_{\rm Hxc}-\left[\frac{\partial \overline{e}_{\rm c}(\nu)}{\partial \nu}
-\frac{\partial \tau^{\mathcal{C}}_{\rm c}(n(\mu),\nu)}{\partial \nu}\right]_{\nu=n(\mu)}.
\end{split}
\ee
The density constraint of Eq.~(\ref{eq:exact_dens_mapping_KS_cluster}) combined with Eq.~(\ref{eq:exact_muimp_expression}) allows for an in-principle-exact evaluation of the Hxc potential $v_{\rm Hxc}$. Most importantly, these two equations can be used for designing an alternative (and self-consistent) embedding strategy on the basis of well-identified density-functional approximations. Indeed, in Ht-DMFET, the second term on the right-hand side of Eq.~(\ref{eq:exact_muimp_expression}) is simply dropped, for simplicity [see Eq.~(\ref{eq:DFA_in_HtDMFET})]. If, in addition, we neglect the Householder kinetic correlation density-bi-functional potential correction $\partial \tau^{\mathcal{C}}_{\rm c}(n,\nu)/\partial \nu$ [last term on the right-hand side of Eq.~(\ref{eq:exact_muimp_expression})], we obtain from Eq.~(\ref{eq:exact_dens_mapping_KS_cluster}) the following self-consistent equation,
\be\label{eq:sc_LPFET_eq}
n_{\rm lattice}^{\rm KS}\left(\mu-\tilde{v}_{\rm Hxc}\right)=n^{\mathcal{C}}\left(\mu-\tilde{v}_{\rm Hxc},\tilde{v}_{\rm Hxc}\right),
\ee
from which an approximation $\tilde{v}_{\rm Hxc}\equiv \tilde{v}_{\rm Hxc}(\mu)$ to the Hxc potential can be determined. Eq.~(\ref{eq:sc_LPFET_eq}) is the second main result of this paper. Since $\tilde{v}_{\rm Hxc}$ is now the to-be-optimized quantity on which the embedding fully relies, we refer to the approach as {\it local potential functional embedding theory} (LPFET), in which the key density-functional approximation that is made reads as
\be\label{eq:LPFET_approx_to_vHxc}
{v}_{\rm Hxc}(n)\underset{\rm LPFET}{\approx}\tilde{\mu}^{\rm imp}(n).
\ee
The approach is graphically summarized in Fig.~\ref{Fig:self-consistent-loop-scheme}.
In order to verify that the first HK theorem~\cite{hktheo} still holds at the LPFET level of approximation, let us assume that two chemical potentials $\mu$ and $\mu+\Delta\mu$ lead to the same density. If so, the converged Hxc potentials should differ by $\tilde{v}_{\rm Hxc}\left(\mu+\Delta\mu\right)-\tilde{v}_{\rm Hxc}(\mu)=\Delta\mu$, so that both calculations give the same KS chemical potential value [see Eq.~(\ref{eq:KS_pot_decomp})]. According to Eqs.~(\ref{eq:sc_LPFET_eq}) and (\ref{eq:LPFET_approx_to_vHxc}), it would imply that two different values of the interacting embedded impurity chemical potential can give the same density, which is impossible~\cite{sekaran2021,senjean2017local}. Therefore,
when convergence is reached in Eq.~(\ref{eq:sc_LPFET_eq}), we can generate an approximate map 
\be
\mu\rightarrow n(\mu)
\underset{\rm LPFET}{\approx}
n_{\rm lattice}^{\rm KS}\left(\mu-\tilde{v}_{\rm Hxc}\right)
=
\expval{\hat{n}_0}_{\Psi^{\mathcal{C}}\left(\mu-\tilde{v}_{\rm Hxc},\tilde{v}_{\rm Hxc}\right)}
,
\ee
and compute approximate per-site energies as follows,
\be\label{eq:per_site_ener_LPFFET}
\frac{E(\mu)}{L}+\mu n(\mu)\underset{\rm{LPFET}}{\approx}\expval{\hat{t}_{01}+\hat{U}_0}_{\Psi^{\mathcal{C}}\left(\mu-\tilde{v}_{\rm Hxc},\tilde{v}_{\rm Hxc}\right)},
\ee
since the approximation in Eq.~(\ref{eq:DFA_in_HtDMFET}) is also used in LPFET, as discussed above.\\

Note that Ht-DMFET and LPFET use the same per-site energy expression [see Eq.~(\ref{eq:per_site_ener_HtDMFET})], which is a functional of the interacting cluster's wave function. In both approaches, the latter and the non-interacting lattice share the same density. Therefore, if the per-site energy is evaluated as a function of the lattice filling $n$, both methods will give exactly the same result. However, different energies will be obtained if they are evaluated as functions of the chemical potential value $\mu$ in the interacting lattice. The reason is that Ht-DMFET and LPFET will give different densities. Indeed, as shown in Sec.~\ref{subsubsec:exact_DFE}, Ht-DMFET can be viewed as an approximation to KS-DFT where the Hxc density-functional potential of Eq.~(\ref{eq:approx_Hxc_pot_HtDMFET}) is employed. As readily seen from Eq.~(\ref{eq:LPFET_approx_to_vHxc}), the LPFET and Ht-DMFET Hxc potentials differ by the Householder kinetic correlation potential (which is neglected in LPFET). If the corresponding KS densities were the same then the Hxc potential, the Householder transformation, and, therefore, the chemical potential on the interacting embedded impurity would be the same, which is impossible according to Eqs.~(\ref{eq:approx_Hxc_pot_HtDMFET}) and (\ref{eq:LPFET_approx_to_vHxc}).\\

\subsection{Comparison with SDE}\label{subsec:sde_comparison}

At this point we should stress that LPFET is very similar to the SDE approach of Mordovina {\it et al.}~\cite{mordovina2019self}. The major difference between SDE and LPFET (in addition to the fact that LPFET has a clear connection with a formally exact density-functional embedding theory based on the Householder transformation) is that no KS construction is made within the cluster. Instead, the Hxc potential is directly updated in the KS lattice, on the basis of the correlated embedded impurity density. This becomes even more clear when rewriting Eq.~(\ref{eq:sc_LPFET_eq}) as follows,
\be\label{eq:LPFET_Hxc_pot_from_dens_cluster}
\tilde{v}_{\rm Hxc}=\mu-\left[n_{\rm lattice}^{\rm KS}\right]^{-1}\left(n^{\mathcal{C}}\left(\mu-\tilde{v}_{\rm Hxc},\tilde{v}_{\rm Hxc}\right)\right),
\ee
where $\left[n_{\rm lattice}^{\rm KS}\right]^{-1}:\,n\rightarrow \mu_{\rm s}(n)$ is the inverse of the non-interacting chemical-potential-density map. A practical advantage of such a procedure (which remains feasible since the full system is treated at the non-interacting KS level only) lies in the fact that the KS construction within the cluster is automatically (and exactly) generated by the Householder transformation, once the density has been updated in the KS lattice (see Eq.~(\ref{eq:occ_embedded_imp_lattice_equal}) and the comment that follows). Most importantly, the density in the KS lattice and the density of the non-interacting KS embedded impurity (which, unlike the embedded {\it interacting} impurity, is not used in the actual calculation) will match {\it at each iteration} of the Hxc potential optimization process, as it should when convergence is reached. If, at a given iteration, the KS construction were made directly within the cluster, there would always be a ``delay'' in density between the KS lattice and the KS cluster, which would only disappear at convergence. Note that, when the latter is reached, the (approximate) Hxc potential of the lattice should match the one extracted from the cluster, which is defined in SDE as the difference between the KS cluster Hamiltonian and the one-electron part of the interacting cluster's Hamiltonian~\cite{mordovina2019self}, both reproducing the density of the KS lattice. Therefore, according to Eqs.~(\ref{eq:dens_func_cluster_hamilt}), (\ref{eq:Tcluster_op_dens_func}) and (\ref{eq:dens_func_cluster_KS_eq}), the converged Hxc potential will simply correspond to the chemical potential on the interacting embedded impurity, exactly like in LPFET [see Eq.~(\ref{eq:LPFET_approx_to_vHxc})].\\

Note finally that the simplest implementation of LPFET, as suggested by Eq.~(\ref{eq:LPFET_Hxc_pot_from_dens_cluster}), can be formally summarized as follows:   
\be\label{eq:LPFET_algo}
\begin{split}
\tilde{v}^{(i+1)}_{\rm Hxc}&=\mu-\left[n_{\rm lattice}^{\rm KS}\right]^{-1}\left(n^{\mathcal{C}}\left(\mu-\tilde{v}^{(i)}_{\rm Hxc},\tilde{v}^{(i)}_{\rm Hxc}\right)\right),
\\
\tilde{v}^{(i=0)}_{\rm Hxc}&=0.
\end{split}
\ee
A complete description of the algorithm is given in the next section.

\section{LPFET algorithm}\label{sec:lpfet_algo}

\begin{figure}[!htbp]
\begin{center}
\includegraphics[scale=0.27]{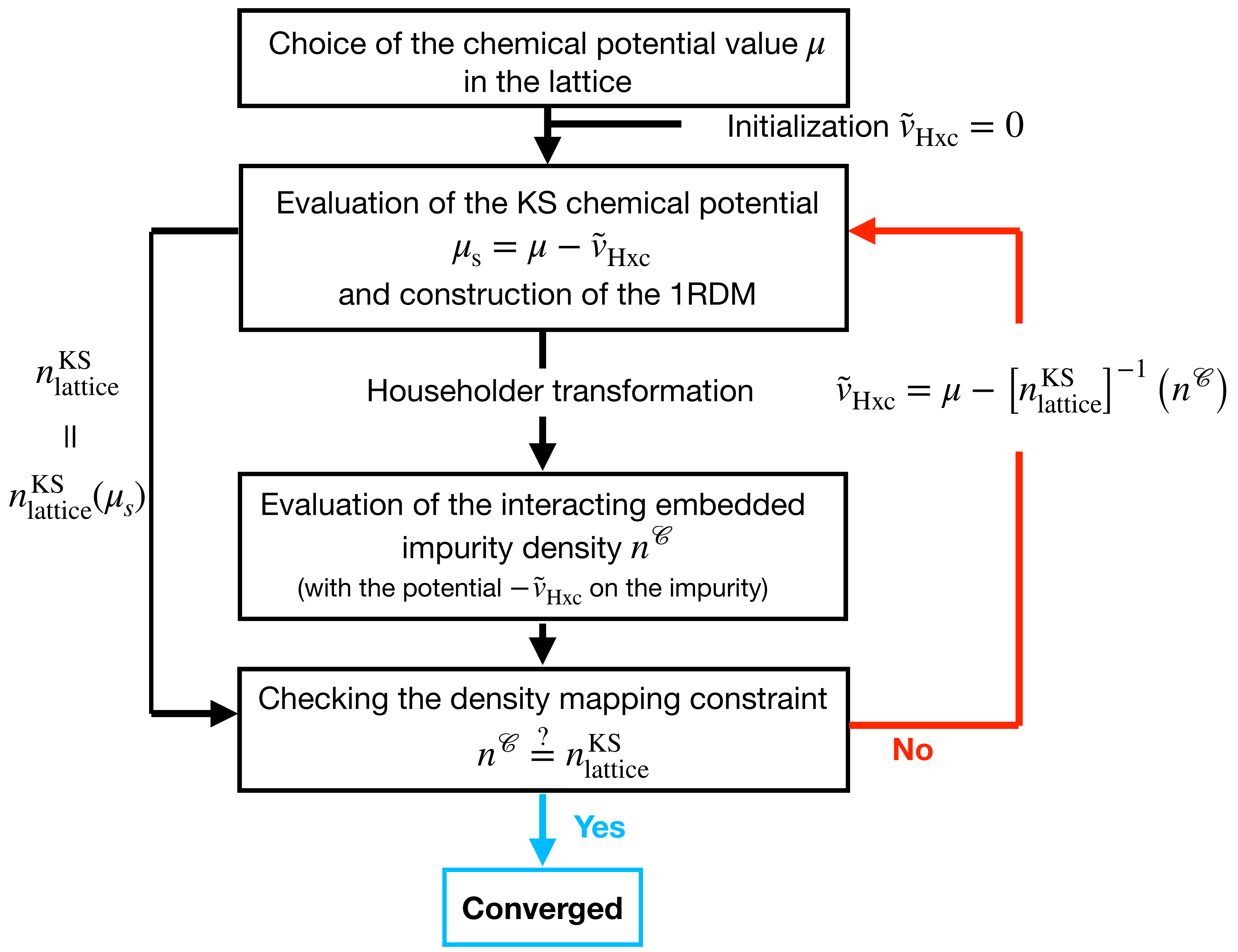}
\caption{Schematics of the LPFET algorithm. The (one-electron reduced) density matrix of the KS lattice is referred to as the 1RDM. See text for further details.}
\label{Fig:self-consistent-loop-convergence}
\end{center}
\end{figure}

The LPFET approach introduced in Sec.~\ref{sec:LPFET} aims at computing the interacting chemical-potential-density $\mu \rightarrow n(\mu)$ map through the self-consistent optimization of the uniform Hxc potential. A schematics of the algorithm is provided in Fig.~\ref{Fig:self-consistent-loop-convergence}. It can be summarized as follows.
\newline
\\1. We start by diagonalizing the one-electron Hamiltonian (\ie, the hopping in the present case) matrix ${\bm t}\equiv t_{ij}$ [see Eq.~(\ref{eq:hopping_matrix})]. Thus we obtain the ``molecular'' spin-orbitals and their corresponding energies. We fix the chemical potential of the interacting lattice to some value $\mu$ and (arbitrarily) initialize the Hxc potential to $\tilde{v}_{\rm Hxc}=0$. Therefore, at the zeroth iteration, the KS chemical potential $\mu_{\rm s}$ equals $\mu$.
\newline
\\
2. We occupy all the molecular spin-orbitals with energies below $\mu_{\rm s}=\mu-\tilde{v}_{\rm Hxc}$ and construct the corresponding density matrix (in the lattice representation). The latter provides the uniform KS density (denoted $n^{\rm KS}_{\rm lattice}$ in Fig.~\ref{Fig:self-consistent-loop-convergence}) and the embedding Householder cluster Hamiltonian [see Eq.~(\ref{eq:int_cluster_SE})] in which the impurity chemical potential is set to $\tilde{\mu}^{\rm imp}=\tilde{v}_{\rm Hxc}$ [see Eq.~(\ref{eq:LPFET_approx_to_vHxc})].    
\newline
\\ 3. We solve the interacting Schr\"{o}dinger equation for the two-electron Householder cluster and deduce the occupation of the  embedded impurity (which is denoted $n^{\mathscr{C}}$ in Fig.~\ref{Fig:self-consistent-loop-convergence}). This can be done analytically since the Householder cluster is an asymmetric Hubbard dimer~\cite{sekaran2021}. 
\newline
\\ 4. We verify that the density in the KS lattice $n^{\rm KS}_{\rm lattice}$ and the occupation of the interacting embedded impurity $n^{\mathscr{C}}$ match (a convergence threshold has been set to 10$^{-4}$). If this is the case, the calculation has converged and $n^{\mathscr{C}}$ is interpreted as (an approximation to) the density $n(\mu)$ in the true interacting lattice. If the two densities do not match, the Hxc potential $\tilde{v}_{\rm Hxc}$ is adjusted in the KS lattice such that the latter reproduces $n^{\mathscr{C}}$ [see Eq.~(\ref{eq:LPFET_algo})] or, equivalently, such that the KS lattice contains $Ln^{\mathscr{C}}$ electrons. We then return to step 2.\\ 

\color{black}

 \section{Results and discussion}\label{sec:results}
In the following, LPFET is applied to a uniform Hubbard ring with a large $L$ = 1000 number of sites in order to approach the thermodynamic limit. Periodic boundary conditions have been used. The hopping parameter is set to $t$ = 1.
As pointed out in Sec.~\ref{sec:LPFET}, plotting the Ht-DMFET (which is equivalent to DMET or DET for a single embedded impurity) and LPFET per-site energies as functions of the lattice filling $n$ would give exactly the same results (we refer the reader to Ref.~\cite{sekaran2021} for a detailed analysis of these results). However, the chemical-potential-density $\mu\rightarrow n(\mu)$ maps obtained with both methods are expected to differ since they rely on different density-functional approximations [see Eqs.~(\ref{eq:approx_Hxc_pot_HtDMFET}) and (\ref{eq:LPFET_approx_to_vHxc})]. We focus on the self-consistent evaluation of the LPFET map in the following. Comparison is made with Ht-DMFET and the exact BA results.\\

\begin{figure}[!htbp]
\begin{center}
\includegraphics[scale=0.6]{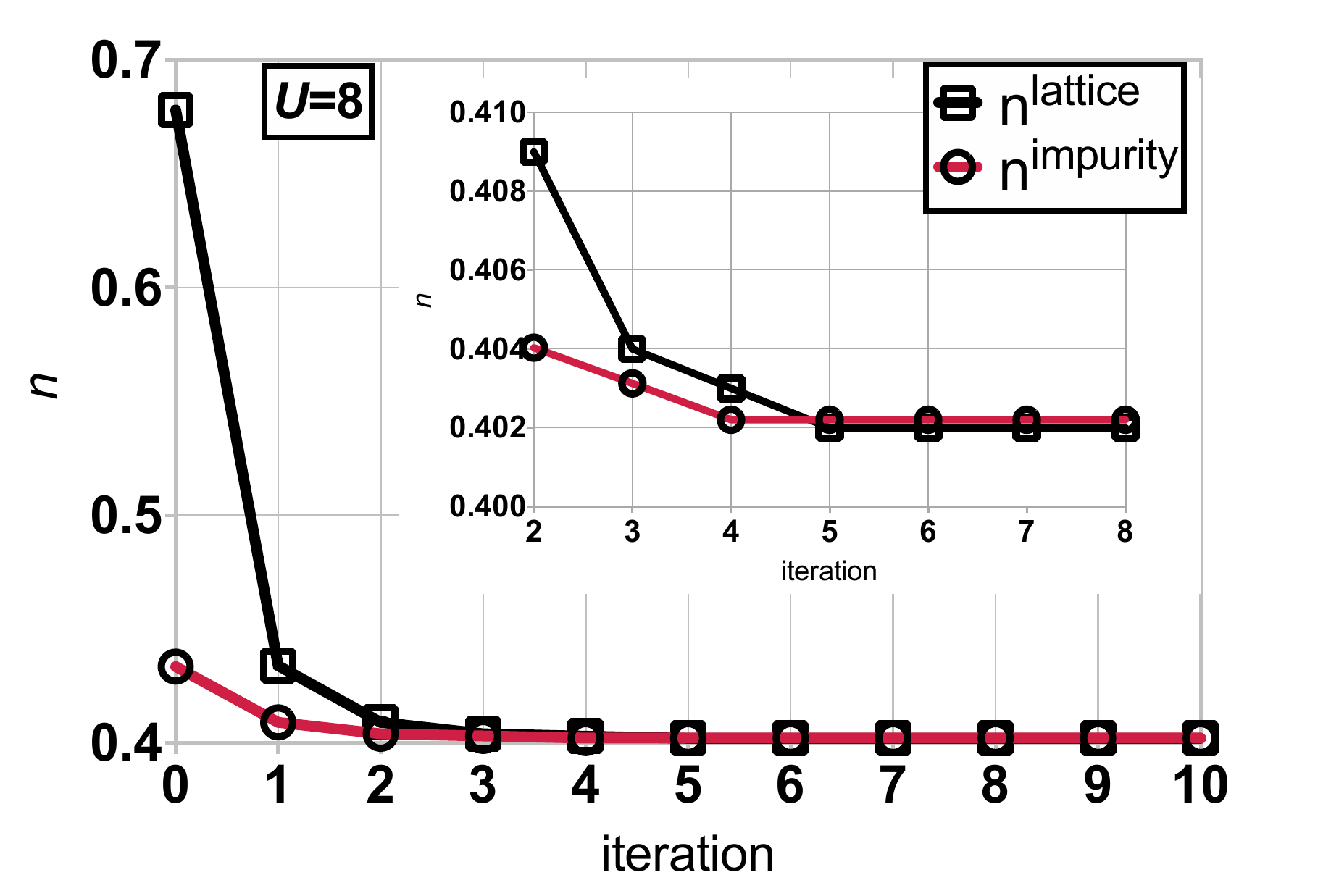}
\caption{Comparison of the KS lattice and embedded impurity densities at each iteration of the LPFET calculation. The interaction strength and chemical potential values are set to $U/t=8$ and $\mu/t = - 0.97$, respectively. As shown in the inset, convergence is reached after five iterations.}
\label{fig:convergence_density}
\end{center}
\end{figure}

\begin{figure}[!htbp]
\begin{center}
\includegraphics[scale=0.6]{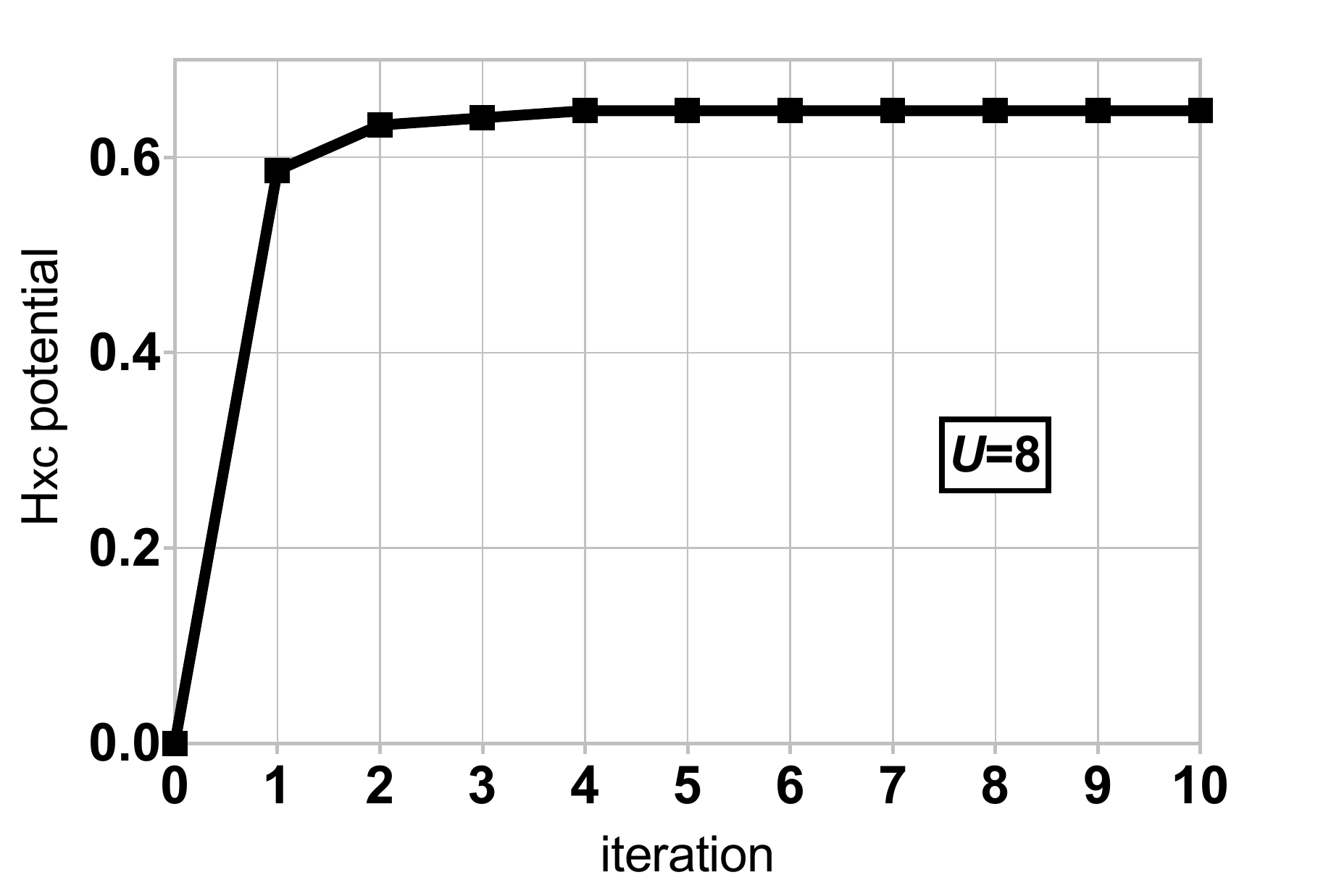}
\caption{Convergence of the LPFET Hxc potential for $U/t=8$ and $\mu/t = - 0.97$.}
\label{fig:convergence_vHxc}
\end{center}
\end{figure}

\begin{figure}[!htbp]
\begin{center}
\includegraphics[scale=0.6]{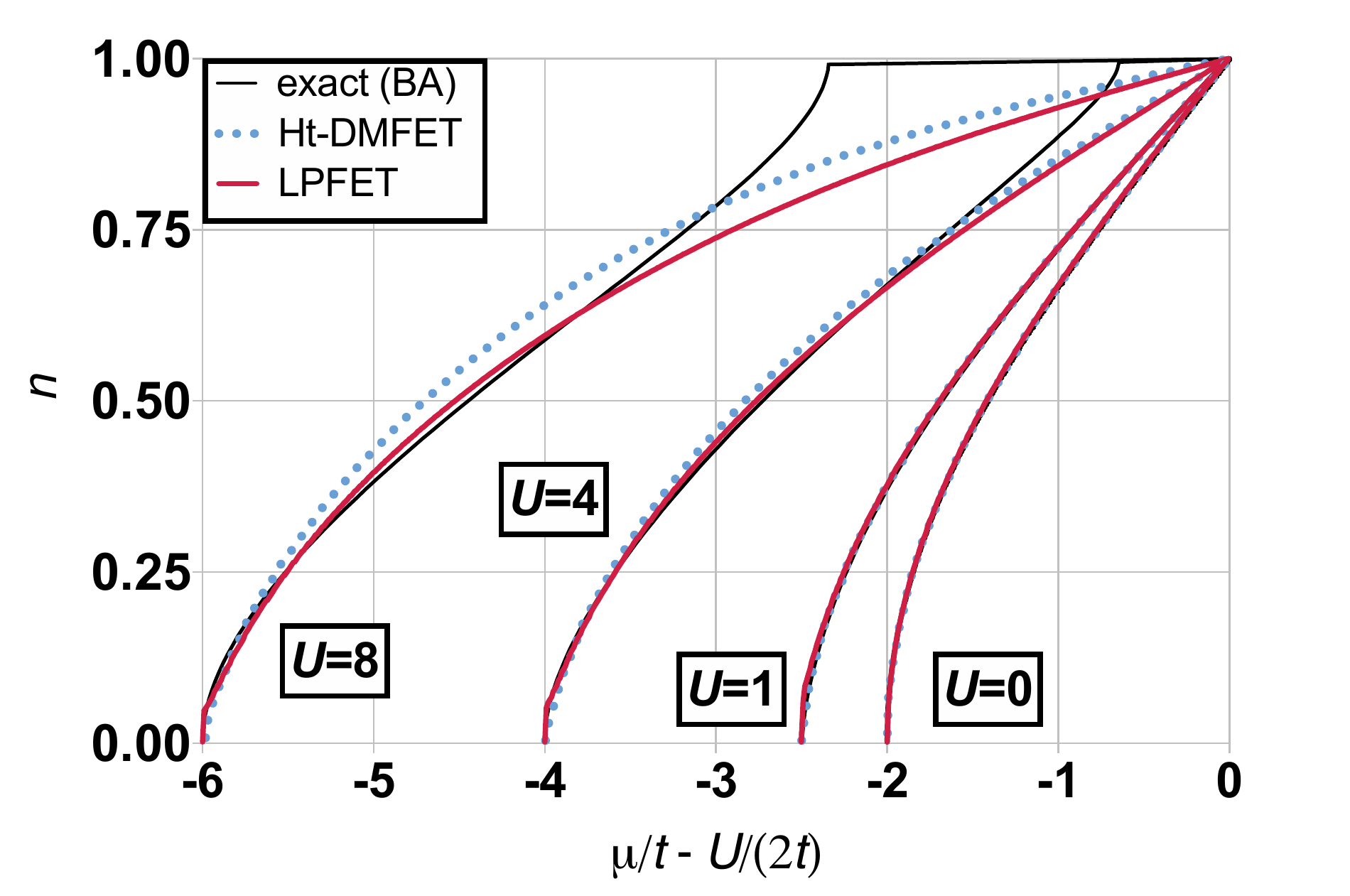}
\caption{Converged LPFET densities (red solid lines) plotted as functions of the chemical potential $\mu$ in various correlation regimes. Comparison is made with the exact BA (black solid lines) and Ht-DMFET (blue dotted lines) results. In the latter case, the chemical potential is evaluated {\it via} the numerical differentiation of the density-functional Ht-DMFET per-site energy [see Eqs.~(\ref{eq:f_cluster_func}) and (\ref{eq:true_chem_pot_Ht-DMFET})]. The non-interacting ($U=0$) chemical-potential-density map [see Eq.~(\ref{eq:KS_dens_func_chemical_potential})] is shown for analysis purposes.} 
\label{fig:Mott-Hubbard}
\end{center}
\end{figure}

\begin{figure}[!htbp]
\begin{center}
\includegraphics[scale=0.6]{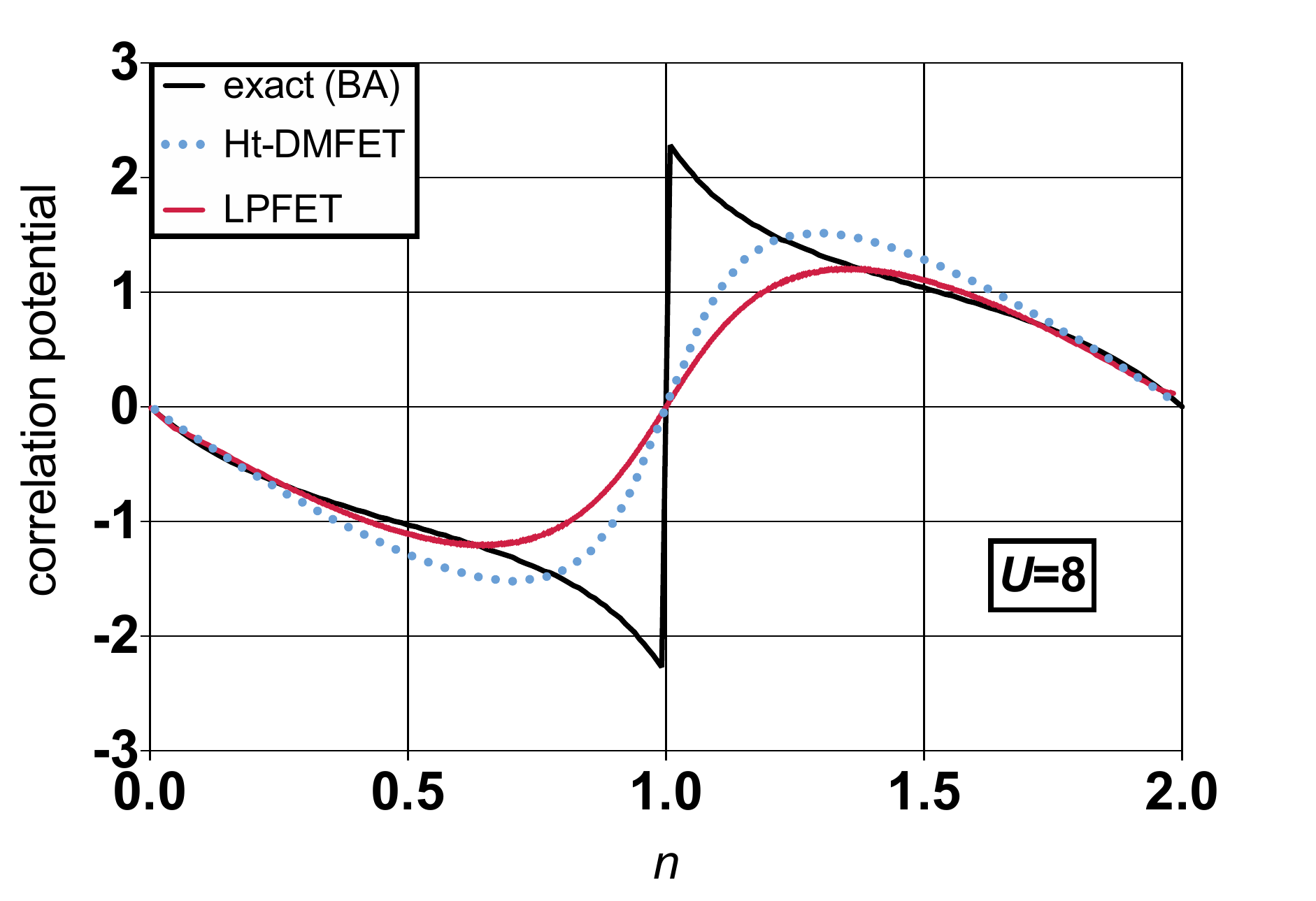}
\caption{Correlation potential $v_{\rm c}(n)=\mu(n)-\mu_{\rm s}(n)-\frac{U}{2}n$ plotted as a function of the lattice filling $n$ at the Ht-DMFET (blue dashed line) and LPFET (red solid line) levels of approximation for $U/t=8$. Comparison is made with the exact BA correlation potential (black solid line).} 
\label{fig:Hartree-exchange-correlation-potential}
\end{center}
\end{figure}

As illustrated by the strongly correlated results of Figs.~\ref{fig:convergence_density} and \ref{fig:convergence_vHxc}, the LPFET self-consistency loop converges smoothly in few iterations. The same observation is made in weaker correlation regimes (not shown). The deviation in density between the KS lattice and the embedded impurity is drastically reduced after the first iteration (see Fig.~\ref{fig:convergence_density}). This is also reflected in the large variation of the Hxc potential from the zeroth to the first iteration (see Fig.~\ref{fig:convergence_vHxc}). It originates from the fact that, at the zeroth iteration, the Hxc potential is set to zero in the lattice while, in the embedding Householder cluster, the interaction on the impurity site is ``turned on''. As shown in Fig.~\ref{fig:convergence_density}, the occupation of the interacting embedded impurity is already at the zeroth iteration a good estimate of the self-consistently converged density. A few additional iterations are needed to refine the result.     
The converged LPFET densities are plotted in Fig.~\ref{fig:Mott-Hubbard} as functions of the chemical potential $\mu$ in various correlation regimes. The non-interacting $U=0$ curve describes the KS lattice at the zeroth iteration of the LPFET calculation. Thus we can visualize, as $U$ deviates from zero, how much the KS lattice learns from the interacting two-electron Householder cluster. LPFET is actually quite accurate (even more than Ht-DMFET, probably because of error cancellations) in the low-density regime. Even though LPFET deviates from Ht-DMFET when electron correlation is strong, as expected, their chemical-potential-density maps are quite similar. This is an indication that neglecting the Householder kinetic correlation potential contribution to the Hxc potential, as done in LPFET, is not a crude approximation, even in the strongly correlated regime. As expected~\cite{knizia2012density,sekaran2021}, LPFET and Ht-DMFET poorly perform when approaching half filling. They are unable to describe the density-driven Mott--Hubbard transition (\ie, the opening of the gap). As discussed in Ref.~\cite{sekaran2021}, this might be related to the fact that, in the exact theory, the Householder cluster is not disconnected from its environment and it contains a fractional number of electrons, away from half filling, unlike in the (approximate) Ht-DMFET and LPFET schemes. In the language of KS-DFT, modeling the gap opening is equivalent to modeling the derivative discontinuity in the density-functional correlation potential $v_{\rm c}(n)=\mu(n)-\mu_{\rm s}(n)-\frac{U}{2}n$ at half filling. As clearly shown in  Fig.~\ref{fig:Hartree-exchange-correlation-potential}, Ht-DMFET and LPFET do not reproduce this feature. In the language of the exact density-functional embedding theory derived in Sec.~\ref{subsec:exact_dfe_dft}, both Ht-DMFET and LPFET approximations neglect the complementary density-functional correlation energy $\overline{e}_{\rm c}(n)$ that is induced by the environment of the (closed) density-functional Householder cluster. As readily seen from Eq.~(\ref{eq:final_vHxc_exp_from_muimp}), it should be possible to describe the density-driven Mott--Hubbard transition with a single statically embedded impurity, provided that we can model the derivative discontinuity in $\partial\overline{e}_{\rm c}(n)/\partial n$ at half filling. This is obviously a challenging task that is usually bypassed by embedding more impurities~\cite{knizia2012density,sekaran2021}. The implementation of a multiple-impurity LPFET as well as its generalization to higher-dimension lattice or quantum chemical Hamiltonians is left for future work.          

\section{Conclusion and perspectives}\label{sec:conclusion}

An in-principle-exact density-functional reformulation of the recently proposed {\it Householder transformed density matrix functional embedding theory} (Ht-DMFET)~\cite{sekaran2021} has been derived for the uniform 1D Hubbard Hamiltonian with a single embedded impurity. On that basis, an approximate {\it local potential functional embedding theory} (LPFET) has been proposed and implemented. Ht-DMFET, which is equivalent to DMET or DET in the particular case of a single impurity, is reinterpreted in this context as an approximation to DFT where the complementary density-functional correlation energy $\overline{e}_{\rm c}(n)$ induced by the environment of the embedding ``impurity+bath'' cluster is neglected. LPFET neglects, in addition, the kinetic correlation effects induced by the Householder transformation on the impurity chemical potential. We have shown that combining the two approximations is equivalent to approximating the latter potential with the Hxc potential of the full lattice. Thus an approximate Hxc potential can be determined {\it self-consistently} for a given choice of external (chemical in the present case) potential in the true interacting lattice. The self-consistency loop, which does not exist in regular single-impurity DMET or DET~\cite{PRB21_Booth_effective_dynamics_static_embedding}, emerges naturally in LPFET from the exact density constraint, \ie, by forcing the KS lattice and interacting embedded impurity densities to match. In this context, the energy becomes a functional of the Hxc potential. In this respect, LPFET can be seen as a flavor of KS-DFT where no density functional is used. LPFET is very similar to SDE~\cite{mordovina2019self}. The two approaches essentially differ in the optimization of the potential. In LPFET, no KS construction is made within the embedding cluster, unlike in SDE. Instead, the Hxc potential is directly updated in the lattice. As a result, the KS cluster (which is not used in the actual calculation) can be automatically generated with the correct density by applying the Householder transformation to the KS lattice Hamiltonian.\\

LPFET and Ht-DMFET chemical-potential-density maps have been computed for a 1000-site Hubbard ring. Noticeable differences appear in the strongly correlated regime. LPFET is more accurate than Ht-DMFET in the low-density regime, probably because of error cancellations. As expected from previous works~\cite{sekaran2021,knizia2012density}, their performance deteriorates as we approach half filling. It appears that, in the language of density-functional embedding theory, it should be possible to describe the density-driven Mott--Hubbard transition (\ie, the opening of the gap), provided that the complementary correlation potential $\partial\overline{e}_{\rm c}(n)/\partial n$ exhibits a derivative discontinuity at half filling. Since the latter is neglected in both methods, the gap opening is not reproduced. The missing correlation effects might be recovered by applying a multi-reference G\"{o}rling--Levy-type perturbation theory on top of the correlated cluster calculation~\cite{sekaran2021}. Extending LPFET to multiple impurities by means of a block Householder transformation is another viable strategy~\cite{sekaran2021}. Note that, like DMET or SDE, LPFET is in principle applicable to quantum chemical Hamiltonians written in a localized molecular orbital basis. Work is currently in progress in these directions.

 \section*{Acknowledgments} 

The authors thank Saad Yalouz (for his comments on the manuscript and many fruitful discussions) and Martin Rafael Gulin (for stimulating discussions). The authors also thank LabEx CSC (ANR-10-LABX-0026-CSC) and ANR
(ANR-19-CE29-0002 DESCARTES and ANR-19-CE07-0024-02 CoLab projects) for funding.



\begin{appendices}
\appendix
\numberwithin{equation}{section}
\setcounter{equation}{0}
\section{Simplification of density matrix elements in the Householder representation}

Starting from the expression in Eq.~(\ref{eq:Householder_creation_ops}) of the creation operators in the Householder representation and Eqs.~(\ref{eq:P_from_HH_vec}), (\ref{eq:v0_zero})-(\ref{eq:HH_vec_compt_i_larger2}), and (\ref{eq:normalization_HH_vec}), we can simplify step by step the expression of the density matrix elements that involve the impurity as follows,     
\be
\begin{split}
\mel{\Phi}{\hat{d}_{j\sigma}^\dagger\hat{d}_{0\sigma}}{\Phi}&=\sum_iP_{ji}\gamma_{i0}
\\
&=\gamma_{j0}-2{\rm v}_j\sum_{i\geq 1}{\rm v}_i\gamma_{i0}
\\
&=\gamma_{j0}-2{\rm v}_j{\rm v}_1\gamma_{10}-2{\rm v}_j\sqrt{2\tilde{\gamma}_{10}\left(\tilde{\gamma}_{10}-\gamma_{10}\right)}\sum_{i\geq 2}{\rm v}^2_i
\\
&=\gamma_{j0}-2{\rm v}_j{\rm v}_1\gamma_{10}-2{\rm v}_j\sqrt{2\tilde{\gamma}_{10}\left(\tilde{\gamma}_{10}-\gamma_{10}\right)}\left(1-{\rm v}^2_1\right)
\\
&=\gamma_{j0}-2{\rm v}_j{\rm v}_1\gamma_{10}-2{\rm v}_j\sqrt{2\tilde{\gamma}_{10}\left(\tilde{\gamma}_{10}-\gamma_{10}\right)}
+2{\rm v}_j{\rm v}^2_1\sqrt{2\tilde{\gamma}_{10}\left(\tilde{\gamma}_{10}-\gamma_{10}\right)}
\\
&=\gamma_{j0}-2{\rm v}_j{\rm v}_1\gamma_{10}-2{\rm v}_j\sqrt{2\tilde{\gamma}_{10}\left(\tilde{\gamma}_{10}-\gamma_{10}\right)}+2{\rm v}_j{\rm v}_1\left(\gamma_{10}-\tilde{\gamma}_{10}\right)
\\
&=\gamma_{j0}-2{\rm v}_j\left({\rm v}_1\tilde{\gamma}_{10}+\sqrt{2\tilde{\gamma}_{10}\left(\tilde{\gamma}_{10}-\gamma_{10}\right)}\right)
\\
&=\gamma_{j0}-2{\rm v}_j\sqrt{2\tilde{\gamma}_{10}\left(\tilde{\gamma}_{10}-\gamma_{10}\right)}\left(1+\dfrac{\tilde{\gamma}_{10}\left(\gamma_{10}-\tilde{\gamma}_{10}\right)}{2\tilde{\gamma}_{10}\left(\tilde{\gamma}_{10}-\gamma_{10}\right)}\right)
\\
&=\gamma_{j0}-{\rm v}_j\sqrt{2\tilde{\gamma}_{10}\left(\tilde{\gamma}_{10}-\gamma_{10}\right)}.
\end{split}
\ee

\end{appendices}




\end{document}